\documentclass[%
 reprint,
 amsmath,amssymb,
 aps,
 pra,
]{revtex4-2}

\usepackage{graphicx}
\usepackage{dcolumn}
\usepackage{bm}

\begin{document}


\title{Photons in a Spherical Cavity}

\author{Thomas B. Bahder}
\affiliation{%
Quantique, LLC, U.S.A.}%

\date{\today}

\begin{abstract}
The iconic problem of photon modes in a spherical cavity has been discussed in the literature; however, conflicting results have been reported~\cite{Heitler,Davydov_QuantumMechanics}. For this reason, the solution of this problem is worked out in detail here, starting with the Maxwell equations and applying boundary conditions at the surface of the bounding perfect conductor. Contrary to the treatments in the literature~\cite{Heitler,Davydov_QuantumMechanics}, the allowed frequencies for photons in the sphere are given by two different conditions, one for electric and one for magnetic multipole photons. After establishing the modes and their allowed frequencies, we write down the second-quantized vector potential in the spherical geometry. Based on these spherical modes, bipartite photon entanglement is investigated showing that there are in-principle 40 different types of entangled photon states. Finally, we include some appendices about photon plane-wave and spherical-wave helicity states, helicity spherical harmonic vectors, and rotation of the helicity states and 3-d vectors using the Wigner $D$-matrix.
\end{abstract}

\maketitle

\maketitle

\section{Introduction}
\label{Introduction}
Classical electrodynamics has a long history~\cite{Jackson-3rdEdition} and many fundamental textbook problems have been solved and published. However, the iconic problem of photon modes in a spherical cavity seems to have escaped detailed solution and publication. Indeed, there are two detailed discussions of this problem in the literature, with differing results~\cite{Heitler,Davydov_QuantumMechanics}. Davydov considers the problem of photon modes in a spherical cavity and he gives one condition for allowed frequencies for both electric and magnetic photon multipoles~\cite{Davydov_QuantumMechanics}. On the other hand, Heitler gives a different condition for these modes~\cite{Heitler}. To clarify the solution of this problem, we work out in-detail the photon modes in a spherical cavity (filled with vacuum) of radius $R$ bounded by a perfect conductor.

Section \ref{MaxwellFieldEquations} rederives the classic Helmholtz equation to be solved inside the sphere. Section \ref{BoundaryConditions} gives the transversality condition for the vector potential and the boundary conditions at the surface of the conductor. Section \ref{VectorPotentialModes} has a calculation of the classical vector potential modes and the energy of these modes. As mentioned above, the boundary conditions lead to two different conditions for the allowed photon frequencies, one for electric and one for magnetic multipoles. Section \ref{Quantization} computes the energy of the field modes in terms of the vector potential, and it is used to quantize the field. Section \ref{AllowedFrequencies} contains a numerical calculation of allowed photon frequencies inside a sphere, for the first few electric and magnetic multipole modes.
Section \ref{TransitionProbabilities} contains a brief discussion of the scaling of allowed transition probabilities based on simple parity selection rules. Section \ref{BipartiteEntangledPhotonStates} investigates bipartite entangled photon states in the sphere, based on the computed spherical modes. Finally, several appendices are included about plane-wave and spherical-wave helicity states, and how these states transform under rotation of coordinates using the Wigner $D$-matrix.

\section{Maxwell Field Equations}
\label{MaxwellFieldEquations}
The fields $\mathbf{E}$ and $\mathbf{B}$ inside the spherical cavity must satisfy the vacuum Maxwell equations (in Gaussian units)~\cite{Jackson-3rdEdition}:

\begin{align}
 \nabla \times \mathbf{E} + \frac{1}{c} \frac{\partial \mathbf{B}}{\partial t} &= 0 & \nabla \cdot \mathbf{E} &= 0 \nonumber \\
 \nabla \times \mathbf{B} - \frac{1}{c}\frac{\partial \mathbf{E}}{\partial t} &= 0 & \nabla \cdot \mathbf{B} &= 0
\label{MaxwellEq}
\end{align}
Assuming a time dependence $\exp(-i \omega t)$ for the fields, the above equations become
\begin{align}
 \nabla \times \mathbf{E} &= ik\mathbf{B} & \nabla \cdot \mathbf{E} &= 0 \nonumber \\
 \nabla \times \mathbf{B} &= - ik\mathbf{E} & \nabla \cdot \mathbf{B} &= 0
\label{MaxwellEqSteadyState1}
\end{align}
where $k=\omega/c$.
There are two equivalent sets of equations~\cite{Jackson-3rdEdition} from Eq.~(\ref{MaxwellEqSteadyState1}): by eliminating the field $\mathbf{E}$, Eqs.~(\ref{MaxwellEqSteadyState1}) become
\begin{align}
 \left( \nabla^2 + k^2 \right)\mathbf{B} &= 0 & \nabla \cdot \mathbf{B} &= 0 \\
 \mathbf{E} &= \frac{i}{k}\nabla \times \mathbf{B}
\end{align}
Alternatively, by eliminating the field $\mathbf{B}$, Eqs.~(\ref{MaxwellEqSteadyState1}) become
\begin{align}
 \left( \nabla^2 + k^2 \right)\mathbf{E} &= 0 & \nabla \cdot \mathbf{E} &= 0 \\
 \mathbf{B} &= - \frac{i}{k}\nabla \times \mathbf{E}
\label{MaxwellEqSteadyState2}
\end{align}
Introducing the vector potential $\mathbf{A}$ by
\begin{equation}
\mathbf{B}= \nabla \times \mathbf{A}
\label{VectorPotentialDef}
\end{equation}
Then, from Eqs.~(\ref{MaxwellEqSteadyState1}), we have
\begin{equation}
 \mathbf{E} = i \frac{\omega}{c} \mathbf{A}
\label{E-field}
\end{equation}
and
\begin{equation}
 \nabla \times \mathbf{E} = i \frac{\omega}{c} \mathbf{B}
\label{E-field2}
\end{equation}

Using the first equation in Eq.~(\ref{MaxwellEqSteadyState1}), taking the curl, and using the third equation in
Eq.~(\ref{MaxwellEqSteadyState1}) for $\nabla \times \mathbf{B}$, we obtain the wave equation satisfied by $\mathbf{E}$:
\begin{equation}
\nabla \times \nabla \times \mathbf{E} = k^2\mathbf{E}
\label{waveEq1}
\end{equation}
Using Eq.~(\ref{E-field}), we get the standard wave equation for the vector potential $\mathbf{A}$:
\begin{equation}
\nabla \times \nabla \times \mathbf{A} = k^2\mathbf{A}
\label{waveEq1_A}
\end{equation}
Applying the vector identity $\nabla \times \nabla \times \mathbf{A} = \nabla (\nabla \cdot \mathbf{A}) - \nabla^2 \mathbf{A} $,
and choosing the Coulomb gauge,
\begin{equation}
\nabla \cdot \mathbf{A} = 0
\label{CoulombGauge}
\end{equation}
we get the well-known vector Helmholtz equation for $\mathbf{A}$:
\begin{equation}
\left( \nabla^2 + k^2 \right)\mathbf{A} = 0
\label{HelmholtzEq}
\end{equation}
In what follows, we write down the solutions of Eq.~(\ref{HelmholtzEq}) for a spherical cavity, bounded by a perfect conductor, subject to the constraint that $\mathbf{A}$ is transverse to the direction of propagation, given in Eq.~(\ref{CoulombGauge}), and subject to boundary conditions on the fields $\mathbf{E}$ and $\mathbf{B}$.

\section{Boundary Conditions}
\label{BoundaryConditions}
The fields $\mathbf{E}$ and $\mathbf{B}$ inside the spherical cavity (which is centered at the origin of spherical coordinates and has radius $R$) must have a finite value at the origin of coordinates $r=0$. This imposes certain constraints on the radial solution, see below. Furthermore, the fields $\mathbf{E}$ and $\mathbf{B}$ must satisfy certain conditions, which are derivable from the field Eqs.~(\ref{MaxwellEq}):
\begin{enumerate}
 \item The vector potential $\mathbf{A}$ must satisfy the vector Helmholtz equation~(\ref{HelmholtzEq}).
 \item The tangential electric field must vanish at the surface of the conductor at $r=R$:
 \begin{align}
 \mathbf{E} \cdot \hat{\mathbf{\theta}}\vert_{r=R} &= 0 \label{BC1} \\
 \mathbf{E} \cdot \hat{\mathbf{\phi}}\vert_{r=R} &= 0 \label{BC2}
 \end{align}
 where $\hat{\mathbf{\theta}}$ and $\hat{\mathbf{\phi}}$ are the usual spherical coordinate unit vectors.
In view of Eq.~(\ref{E-field}), this translates to vanishing of the components of the vector potential at the surface of the conductor:
\begin{align}
 \mathbf{A} \cdot \hat{\theta}\vert_{r=R} &= 0 \label{ABC1} \\
 \mathbf{A} \cdot \hat{\phi}\vert_{r=R} &= 0 \label{ABC2}
 \end{align}
 \item The normal magnetic field must vanish at the surface of the conductor:
\begin{equation}
 \mathbf{B} \cdot \mathbf{n}\vert_{r=R} = 0
\label{BFieldNormalBC}
\end{equation}
where $\mathbf{n}$ is the normal to the spherical surface of the cavity.
 In view of Eq.~(\ref{VectorPotentialDef}), this imposes the condition on the vector potential
 \begin{equation}
 \mathbf{n} \cdot \nabla \times \mathbf{A}\vert_{r=R} = 0
\label{AFieldNormalBC}
\end{equation}
\item Coulomb gauge condition, given by Eq.~(\ref{CoulombGauge}), must be satisfied inside the sphere of radius $R$ by the vector potential $\mathbf{A}$. At each point inside the sphere, this condition makes the $\mathbf{E}$-field normal to the direction of propagation $\mathbf{k}$, which makes it a transverse field. In terms of the vector potential in $\mathbf{k}$-space, this condition is
\begin{equation}
\mathbf{k} \cdot \mathbf{A}(\mathbf{k}) = 0
\label{transversK}
\end{equation}
\end{enumerate}

\section{Vector Potential Modes}
\label{VectorPotentialModes}
The solutions of the vector Helmholtz Eq.~(\ref{HelmholtzEq}) are well-known for spherical geometry~\cite{Jackson-3rdEdition}. The solutions are given by a product of a radial function, $R_l(r)$, and spherical vector harmonics $\mathbf{Y}_{j l m}$:
\begin{equation}
 \mathbf{A}(\mathbf{r}) \sim R_l(r) \mathbf{Y}_{jlm}(\theta,\phi)
\label{HelmholtzSolution}
\end{equation}
The vector spherical harmonics $\mathbf{Y}_{j l m}$ are eigenfunctions of the square of the total angular momentum, $\hat{\mathbf{J}}^2$, the square of the orbital angular momentum squared, $\hat{\mathbf{L}}^2$, the square of the spin angular momentum squared, $\hat{\mathbf{S}}^2$, and the z-component of total angular momentum, $J_z$, with corresponding eigenvalues: $j(j+1)$, $l(l+1)$, $s(s+1)=2$, and $m$, respectively, see Appendix~\ref{VectorSphericalHarmonics} and \ref{VectorSphericalHarmonicsHelicityFunctions}. There are three types of spherical harmonic functions $\mathbf{Y}_{jlm}\left( {\theta ,\phi } \right)$ for the three possible values of $l=j-1,j,j+1$.

Note that, in order for $\mathbf{A}(\mathbf{r})$ to be a solution of Eq.~(\ref{HelmholtzEq}), the orbital angular momentum $l$ must be the same for the radial function $R_l(r)$ and the spherical harmonic $\mathbf{Y}_{j,l,m}(\theta,\phi)$. This will become an issue when writing down the electric multipole solution for the electric case, see below. In what follows, we review the solution of the boundary value problem inside the sphere~\cite{Jackson-3rdEdition}.

The vector Helmholtz Eq.~(\ref{HelmholtzEq}) can be written as
\begin{equation}
\left[ {\frac{1}{r^2}\frac{\partial^2}{\partial r^2}r - \frac{\hat{L}^2}{r^2} + k^2} \right]\mathbf{A}(\mathbf{r}) = 0
\label{HelmholtzEq2}
\end{equation}
where the square of the orbital angular momentum operator is given by
\begin{equation}
\hat{L}^2 = - \left[ \frac{1}{\sin \theta}\frac{\partial}{\partial \theta}\left( \sin \theta \frac{\partial}{\partial \theta} \right) + \frac{1}{\sin^2 \theta}\frac{\partial^2}{\partial \phi^2} \right]
\label{OrbitalAMOperator}
\end{equation}
and the spherical (scalar) harmonics $Y_{lm}(\theta,\phi)$ satisfy the well-known eigenvalue equation~\cite{Jackson-3rdEdition}
\begin{equation}
\hat{L}^2 \,Y_{lm} = l(l + 1)Y_{lm}
\label{SphericalHarmonicsEigenvalueEq}
\end{equation}
In this work, we use the phase convention of Landau and Lifshitz for the spherical harmonic functions:~\cite{LL_QuantumMechanics,LL_QuantumElectrodynamics}

\begin{align}
Y_{lm}(\theta ,\phi) = \,\, i^l \, (-1)^{(m + |m|)/2} \, & \left[ \frac{(2l + 1)\,(l - |m|)!}{4\pi (l + |m|)!} \right]^{1/2} \nonumber \\
 & \times P_l^{|m|}(\cos \theta)e^{im\phi}
\end{align}
where $P_l^{|m|}(\cos \theta)$ are the associated Legendre functions.

Substituting the form of the vector potential in Eq.~(\ref{HelmholtzSolution}) into Eq.~(\ref{HelmholtzEq2}), and canceling terms, we get the ordinary differential equation satisfied by the radial function $R_l(r)$:
\begin{equation}
\frac{d^2 R_l}{dr^2} + \frac{2}{r}\frac{dR_l}{dr} + \left( k^2 - \frac{l(l + 1)}{r^2} \right)R_l(r) = 0
\label{radialEq}
\end{equation}
Substituting
\begin{equation}
 R_l(r)= \frac{1}{r^{1/2}} u_l(r)
\label{radialEq2}
\end{equation}

\begin{equation}
\left[ \frac{d^2}{dr^2} + \frac{1}{r}\frac{d}{dr} + k^2 - \frac{(l + \frac{1}{2})^2}{r^2} \right]u_l(r) = 0
\label{radialEq3}
\end{equation}
Further, substituting $x=k r$ and $u_l(r)=w(k r)$ into the above, we get the well-known Bessel differential equation of order $\nu= l+\frac{1}{2}$:
\begin{equation}
\frac{d^2w(x)}{dx^2} + \frac{1}{x}\frac{dw(x)}{dx} + \left( 1 - \frac{\nu^2}{x^2} \right)w(x) = 0
\label{BesselEq}
\end{equation}
The regular solutions at $r=0$ of Eq.~(\ref{BesselEq}) are Bessel functions of the first kind, $J_\nu(x)$~\cite{Jackson-3rdEdition}
\begin{equation}
R_l(r) = \frac{u_l(r)}{r^{1/2}} = \frac{J_{l + \frac{1}{2}}(kr)}{r^{1/2}}
\label{functionDefs}
\end{equation}
As mentioned above the field solutions must be finite at the origin $r=0$. Therefore, the radial solutions $R_{l}(r)$ of Eq.~(\ref{radialEq}) are, up to a multiplicative constant, the spherical Bessel functions of integer order $l$ defined by
\begin{equation}
j_l(kr) = \left( \frac{\pi}{2kr} \right)^{1/2} J_{l + \frac{1}{2}}(k r)
\label{sphericalBesselFn}
\end{equation}
where $J_{l + \frac{1}{2}}(kr)$ is the Bessel function of the first kind~\cite{Jackson-3rdEdition}.
The vector potential $\mathbf{A}$ must be transverse, i.e., it must satisfy Eq.~(\ref{CoulombGauge}). Generally, the vector spherical harmonics (VSH) $\mathbf{Y}_{j l m}(\theta, \phi)$ do not satisfy $\nabla \cdot (R_{l}(r) \mathbf{Y}_{j l m}(\theta, \phi) ) = 0 $. However, linear combinations of these functions can be constructed to do so. In order to proceed, we look at the problem in momentum space, that is, we Fourier transform Eq.(\ref{CoulombGauge}), which leads to the transverse constraint in $\mathbf{k}$-space given by Eq.~(\ref{transversK}).

Considering Eq.(\ref{transversK}) in k-space, we look for linear combinations of vector spherical harmonics (VSH) that satisfy the transversality condition in $\mathbf{k}$-space that satisfy $\mathbf{k} \cdot \mathbf{Y}=0$. This problem has been discussed in Ref.~\cite{LL_QuantumElectrodynamics}. There are three independent spherical harmonics: $\mathbf{Y}_{jm}^{L}( \mathbf{\hat{k}} )$, $\mathbf{Y}_{jm}^{M} ( \mathbf{\hat{k}} )$, and $\mathbf{Y}_{jm}^{E} ( \mathbf{\hat{k}} )$, see Appendix~\ref{VectorSphericalHarmonics}.
The vector $\mathbf{Y}_{jm}^{L}( \mathbf{\hat{k}} )$ is longitudinal, meaning that
$\mathbf{k} \cdot \mathbf{Y}_{jm}^{L}( \mathbf{\hat{k}} ) \ne 0$, so it cannot be used to define a transverse vector potential field~\cite{LL_QuantumElectrodynamics}. The other two vectors satisfy
\begin{equation}
\begin{aligned}
 \mathbf{k} \cdot \mathbf{Y}_{jm}^{M}(\mathbf{k}) &= 0 \\
 \mathbf{k} \cdot \mathbf{Y}_{jm}^{E}(\mathbf{k}) &= 0
\end{aligned}
\label{TransverseY}
\end{equation}
and can be used as a basis for transverse fields, see Appendix~\ref{VectorSphericalHarmonics}. We apply boundary conditons to fields, according to Eq.~(\ref{BC1})-(\ref{AFieldNormalBC}), so the allowed frequencies (or equivalently wave vectors) are discrete. The wave functions in k-space are given by~Ref.~\cite{LL_QuantumElectrodynamics}
\begin{equation}
\mathbf{A}_{\omega jm}^\tau (\mathbf{k}) = C_{\omega jm}^\tau \, \delta_{k,\omega /c} \, \mathbf{Y}_{jm}^{\tau}(\mathbf{\hat{k}})
\label{k-spaceVectorWF}
\end{equation}
where $C_{\omega jm}^\lambda \,$ are normalization constants. The $\delta$-function is a Kronecker delta, which depends on the radial $\mathbf{k}$ value, $k=|\mathbf{k}| $, which imposes energy conservation, so the discrete frequencies $\omega$ (to be determined below) are $ \omega = c k$. The index $\tau =M$ or $E$, specifies the magnetic multipole modes, or electric multipole modes, respectively~\cite{LL_QuantumElectrodynamics}.

We \textit{define} the $\mathbf{k}$-space vector potential normalization as an integral:
\begin{multline}
 \int d^3 k \,\, \left( \mathbf{A}_{\omega 'j'm'}^{\tau '}(\mathbf{k}) \right)^* \cdot \mathbf{A}_{\omega jm}^\tau (\mathbf{k}) \\
 = \delta_{\omega ,\omega '}\delta_{\tau ,\tau '}\delta_{j,j'}\delta_{m,m'}
\label{Knormalization1}
\end{multline}
To evaluate the integral on the left side of Eq.~(\ref{Knormalization1}), the propagation vector can be written as $\mathbf{k} =(k, \hat{\mathbf{k}})$, where the radial $k$ is discrete (as is $\omega$), due to the boundary conditions , while the angular part, $\hat{\mathbf{k}}$, is continuous. To carry out the integrals in Eq.~(\ref{Knormalization1}), we change the integration to spherical coordinates, $d^3k=k^2 dk \, d\Omega_k $, and then change the radial integral over $k$ to a discrete sum, using the formula
\begin{equation}
\int \frac{d^3k}{(2\pi)^3} F(\mathbf{k}) \to \delta \sum\limits_k k^2 \int d\Omega_k \,F(\mathbf{k})
\label{IntegralToSum}
\end{equation}
for arbitrary well-behaved functions $F(\mathbf{k})$, where $\delta=\pi/R$ is the cell size of radial $k$'s in $k$-space .
Here, we use the fact that the allowed radial $k$ are given by roots of spherical Bessel functions, and combinations of spherical Bessel functions, which for large arguments behaves as
\begin{equation}
j_l(kR) \to \frac{1}{kR}\sin \left( kR - \frac{l\pi}{2} \right)
\label{BesselAsymptotic}
\end{equation}
The spacing between radial $k_n$ values is the cell size in k-space
\begin{equation}
k_{n + 1,l} - k_{n,l} = \Delta k = \frac{\pi}{R} \equiv \delta
\label{k-Cell}
\end{equation}
where $j(k_n \, R)=0$ gives the values of radial $k$'s.

Using Eq.~(\ref{IntegralToSum}) and orthogonality of the vector spherical harmonics, we obtain the normalized vector potential from Eq.~(\ref{Knormalization1})
\begin{equation}
\mathbf{A}_{\omega jm}^\tau (\mathbf{k}) = \frac{c}{\omega \sqrt{\delta}} \,\, \delta_{k,\omega /c} \, \mathbf{Y}_{jm}^{\tau}(\mathbf{\hat{k}})
\label{kNormalizedA}
\end{equation}
where the normalization constant is given by~\footnote{The normalization condition defined in
Eq.~(\ref{Knormalization1}) is such that the integral in Eq.~(\ref{Knormalization1}) is defined by Eq.~(\ref{IntegralToSum}). The cell size $\delta$ is essentially the same for electric and magnetic modes, see Eq.~(\ref{AllowedMagneticFrequencies}) and (\ref{AllowedElectricFrequencies}).}
\begin{equation}
|C_{\omega jm}^\tau |=\frac{c}{\omega \sqrt{\delta}}
\end{equation}

Next , we take the inverse Fourier transform of Eq.~(\ref{kNormalizedA}) to obtain the real space vector potential:
\begin{equation}
\mathbf{A}_{\omega jm}^\tau (\mathbf{r}) = \int \frac{d^3k}{(2\pi)^3} \mathbf{A}_{\omega jm}^\tau (\mathbf{k}) e^{i\mathbf{k} \cdot \mathbf{r}}
\label{FourierTransform}
\end{equation}
To carry out this integration over $d^3 k$, we use the same substitution for changing the radial integral to a sum, as given in Eq.(\ref{Knormalization1}). Also, we must replace the discrete Kronecker $\delta$-function, $\delta_{k,\omega/c} \rightarrow \delta \times \delta(k-\omega/c)$, where $\delta$ is the radial k-spacing given in Eq.~(\ref{k-Cell}). The integration for the cases $\tau=M$ and $\tau=E$ must be treated separately, resulting in:
\begin{equation}
\mathbf{A}_{\omega jm}^{M}(\mathbf{r}) = \frac{\omega \sqrt{\delta}}{2\pi^2 c} i^j \,\, j_j\left( \frac{\omega}{c}r \right) \mathbf{Y}_{jm}^M(\theta ,\phi)
\label{MagneticVectorPotentialKNormalized}
\end{equation}
\begin{align}
\mathbf{A}_{\omega jm}^{E}(\mathbf{r}) &= \frac{\omega \sqrt{\delta}}{2\pi^2 c} \frac{i^{j + 1}}{\sqrt{2j + 1}} \left[ \sqrt{j} \, j_{j + 1}(\frac{\omega}{c} r) \mathbf{Y}_{j,j + 1,m}(\theta ,\phi) \right. \nonumber \\
 & \left. - \sqrt{j + 1} \, j_{j - 1}(\frac{\omega}{c} r)\mathbf{Y}_{j,j - 1,m}(\theta ,\phi ) \right]
\label{ElectricVectorPotentialKNormalized}
\end{align}

The real-space magnetic multipole potential $\mathbf{A}_{\omega jm}^{M}(\mathbf{r})$, and electric multipole potential $\mathbf{A}_{kjm}^{E}(\mathbf{r})$, in Eq.(\ref{MagneticVectorPotentialKNormalized}) and (\ref{ElectricVectorPotentialKNormalized}), are normalized according to Eq.(\ref{Knormalization1}),
that is, the Fourier transforms of these real-space vector potentials, in Eq.~(\ref{MagneticVectorPotentialKNormalized}) and (\ref{ElectricVectorPotentialKNormalized}), are normalized according to Eq~(\ref{Knormalization1}).

Up to this point, we have used Gaussian units. In what follows, we transition to SI units, so the results may be conveniently expressed for possible comparison with experiments.

\section{Normalization of Vector Potential and Quantization of the Field}
\label{Quantization}

Consider a field that is periodic in time described by a real-valued field $\mathbf{E}_R(\mathbf{r},t)$. Alternatively, the same real field can be written as the real part of a complex field by $\mathbf{E}_R(\mathbf{r},t) = \text{Re} \,\, \mathbf{E}_C(\mathbf{r}) e^{-i \omega t}$. The time cycle average of the square of the real field is given by $\langle \mathbf{E}_R^2\rangle $. Written in terms of the complex-valued function we have
\begin{equation}
\langle \mathbf{E}_R^2\rangle = \frac{1}{2}\langle \mathbf{E}_C^* \cdot \mathbf{E}_C\rangle
\label{Real-ComplexFields}
\end{equation}

In SI units, the classical energy of the electromagnetic field inside the sphere is given by
\begin{align}
 \mathcal{E}_{Field} &= \frac{1}{2}\int\limits_V \left[ \epsilon_0\mathbf{E}_R^2 + \frac{1}{\mu _0}\mathbf{B}_R^2 \right] \,d^3\mathbf{r} \nonumber \\
 &= \frac{1}{4}\int\limits_V \left[ \epsilon_0\mathbf{E}_C \cdot \mathbf{E}_C^* + \frac{1}{\mu _0}\mathbf{B}_C \cdot \mathbf{B}_C^* \right] \,d^3\mathbf{r}
\label{FieldEnergy}
\end{align}
where $\epsilon_0$ and $\mu_0$ are the permittivity and permeability of the vacuum.

Using Eq.~(\ref{VectorPotentialDef}) and Eq.~(\ref{E-field}), for each vector potential mode (in SI units), the corresponding fields are:
\begin{align}
 \mathbf{E}_{\omega jm}^\tau (\mathbf{r},t) &= i\omega \mathbf{A}_{\omega jm}^\tau (\mathbf{r},t)\label{Emode} \\
 \mathbf{B}_{\omega jm}^\tau (\mathbf{r},t) &= \nabla \times \mathbf{A}_{\omega jm}^\tau (\mathbf{r},t)
 \label{Bmode}
 \end{align}
where $\mathbf{E}_{\omega jm}^\tau (\mathbf{r},t)$ and $\mathbf{B}_{\omega jm}^\tau (\mathbf{r},t)$ are complex fields.

Using these complex fields, the energy in each mode of the field can be written as
\begin{equation}
\mathcal{E}_{\omega jm}^\tau = \frac{1}{4}\int\limits_V \left[ \varepsilon_0\mathbf{E}_{\omega jm}^\tau \cdot \mathbf{E}_{\omega jm}^{\tau \; * } + \frac{1}{\mu_0}\mathbf{B}_{\omega jm}^\tau \cdot \mathbf{B}_{\omega jm}^{\tau \; * } \right] \,d^3\mathbf{r} = \hbar \omega
\label{ModeEnergy}
\end{equation} 
In Eq.~(\ref{ModeEnergy}), we equated the energy of the field mode, $\mathcal{E}_{\omega jm}^\tau$, to the energy of one photon in that mode, $\hbar \omega$. Using the electric mode and magnetic mode fields in terms of the vector potential, given in Eqs.~(\ref{Emode}) and (\ref{Bmode}), Eq.~(\ref{ModeEnergy}) becomes
\begin{equation}
\mathcal{E}_{\omega jm}^\tau = \frac{1}{4}\int\limits_V \left[ \omega^2 \varepsilon_0 |\mathbf{A}_{\omega jm}^\tau|^2 + \frac{1}{\mu_0} |\nabla \times \mathbf{A}_{\omega jm}^\tau|^2 \right] \,d^3\mathbf{r} = \hbar \omega
\label{ModeEnergy2}
\end{equation}
The first term in Eq.~(\ref{ModeEnergy2}) is from the electric component of the field, and the second term is from the magnetic component of the field. It is well-known that the electric and magnetic fields contribute equal amounts to the total energy of the field. Therefore, we can write
\begin{equation}
\mathcal{E}_{\omega jm}^\tau = \frac{1}{2}\omega^2 \varepsilon_0 \int\limits_V |\mathbf{A}_{\omega jm}^\tau|^2 \,d^3\mathbf{r} = \hbar \omega
\label{ModeEnergy3}
\end{equation}
where the total field energy of the mode is twice that of the energy of the electric component of the field.

Next, we write the magnetic and electric multipole modes, in Eq.~(\ref{MagneticVectorPotentialKNormalized}) and in Eq.~(\ref{ElectricVectorPotentialKNormalized}), with arbitrary constants, $\bar{C}_{\omega jm}^M$ and $\bar{C}_{\omega jm}^E$, respectively:
\begin{equation}
\mathbf{A}_{\omega jm}^{M}(\mathbf{r}) = \bar{C}_{\omega jm}^M \,\, j_j\left( \frac{\omega}{c}r \right) \mathbf{Y}_{jm}^M(\theta ,\phi)
\label{MagneticVectorPotentialKNormalized2}
\end{equation}
\begin{align}
\mathbf{A}_{\omega j m}^{E}(\mathbf{r}) &= \bar{C}_{\omega jm}^E \, \left[ \sqrt{j} \, j_{j + 1} \left(\frac{\omega}{c} r \right) \mathbf{Y}_{j,j + 1,m}(\theta ,\phi ) \right. \nonumber \\
 & \left. - \sqrt{j + 1} \, j_{j - 1} \left(\frac{\omega}{c} r \right)\mathbf{Y}_{j,j - 1,m}(\theta ,\phi ) \right]
\label{ElectricVectorPotentialKNormalized2}
\end{align}

As remarked earlier, the allowed frequencies $\omega$ in
Eq.~(\ref{MagneticVectorPotentialKNormalized2}) and (\ref{ElectricVectorPotentialKNormalized2}) are discrete and they are different for magnetic multipoles and electric multipoles. We label the allowed frequencies as $\omega _{j,n}^M$ and $\omega _{j,n}^E$ in
Eqs.~(\ref{MagneticVectorPotentialKNormalized2}) and (\ref{ElectricVectorPotentialKNormalized2}), for magnetic and electric modes. The constants, $\bar{C}_{\omega jm}^M$ and $\bar{C}_{\omega jm}^E$, will be determined by equating the energy in each mode, $M$ or $E$, to the one-photon energy $\hbar \omega$ given in Eq.~(\ref{ModeEnergy3}). Next, we substitute in turn each of these multipole modes, in Eq.~(\ref{MagneticVectorPotentialKNormalized2}) and (\ref{ElectricVectorPotentialKNormalized2}), into
Eq.~(\ref{ModeEnergy3}), to determine the normalization constants $\bar{C}_{\omega jm}^M$ and $\bar{C}_{\omega jm}^E$:
\begin{equation}
\bar{C}_{\omega jm}^M = \left( \frac{8\hbar}{\pi \epsilon_0 c} \right)^{1/2} \frac{1}{R} \,
 \frac{1}{| J_{j + \frac{3}{2}} \left( \frac{\omega _{j,n}^M}{c} R \right) |}
\label{NormalizationMagnetic}
\end{equation}

\begin{align}
\bar{C}_{\omega jm}^E &= \left( \frac{8\hbar}{\pi \epsilon_0 c} \right)^{1/2} \frac{1}{R} \, \left\{ j \left[ J_{j + \frac{5}{2}}\left( \frac{\omega _{j,n}^E}{c}R \right) \right]^2 + \right. \nonumber \\
 & \left. (j + 1) \left[ J_{j + \frac{1}{2}} \left( \frac{\omega _{j,n}^E}{c} R \right) \right]^2 \right\}^{-\frac{1}{2}}
\label{NormalizationElectric}
\end{align}
where $J_j(x)$ are the Bessel functions of the first kind of order $j$~\cite{Jackson-3rdEdition}. For each mode, the discrete frequencies, $\omega _{j,n}^M$ and $\omega _{j,n}^E$, are determined by the boundary conditions in Eq.~(\ref{ABC1})-(\ref{AFieldNormalBC}), see below. The index $n$ labels the sequence of allowed frequencies, $\omega _{j,n}^E$ and $\omega _{j,n}^M$, for electric and magnetic modes.

By construction, the magnetic and electric vector potential functions in
Eq.~(\ref{MagneticVectorPotentialKNormalized2}) and (\ref{ElectricVectorPotentialKNormalized2}), are now in SI units of volt-sec/meter.

The allowed frequencies for magnetic and electric modes, ${\omega _{j,n}^M}$ and ${\omega _{j,n}^E}$, are given by imposing the boundary conditions in
Eq.~(\ref{BC1})-(\ref{AFieldNormalBC}) on the modes in
Eq.~(\ref{MagneticVectorPotentialKNormalized2}) and (\ref{ElectricVectorPotentialKNormalized2}), leading to:

\begin{equation}
J_{j + \frac{1}{2}}\left( \frac{\omega _{j,n}^M}{c}R \right) = 0
\label{AllowedMagneticFrequencies}
\end{equation}

\begin{equation}
j\,J_{j + \frac{3}{2}}\left( \frac{\omega _{j,n}^E}{c}R \right) - (j + 1)J_{j - \frac{1}{2}}\left( \frac{\omega _{j,n}^E}{c}R \right) = 0
\label{AllowedElectricFrequencies}
\end{equation}

Here, $\omega _{j,n}^M$ and $\omega _{j,n}^E$ are the allowed frequencies for the magnetic and electric vector potential modes in Eqs.~(\ref{MagneticVectorPotentialKNormalized2}) and
(\ref{ElectricVectorPotentialKNormalized2}), respectively. For a given angular momentum $j$, and either magnetic mode $\tau=M$ or electric mode $\tau=E$, the frequencies are labeled by index $n=1,2,3\dots$, which labels the sequence of roots given by Eq.~(\ref{AllowedMagneticFrequencies}) or (\ref{AllowedElectricFrequencies}). Note that the frequencies $\omega _{j,n}^\tau$ do not depend on the eigenvalue $m$ of operator $\hat{J}_z$, so these states are degenerate in energy. The operator $\hat{J}_z$ is the operator that generates infinitesimal rotations about the z-axis, so this energy degeneracy means that there is rotational symmetry about the z-axis. As mentioned above, the quantity $R$ is the radius of the cavity, which contains the vacuum and is bounded by a perfect conductor.

Equation~(\ref{AllowedMagneticFrequencies}) was given by Davydov~\cite{Davydov_QuantumMechanics} for the allowed frequencies (or allowed wave vectors) for both electric and magnetic modes of the vector potential. On the other hand, Heitler gives Eq.~(\ref{AllowedElectricFrequencies}) for the allowed frequencies of the electric multipole potential~\cite{Heitler}. From explicit calculations, we conclude that there are two different conditions: one for allowed frequencies for the magnetic ($\tau=M$) modes and one for allowed frequencies of the electric ($\tau=E$) modes, given by Eq.~(\ref{AllowedMagneticFrequencies}) and (\ref{AllowedElectricFrequencies}), respectively.

Using the modes in Eq.~(\ref{MagneticVectorPotentialKNormalized2}) and (\ref{ElectricVectorPotentialKNormalized2}), we can write down the classical mode expansion of the vector potential in electric and magnetic multipoles as~\cite{Davydov_QuantumMechanics}:
\begin{equation}
\mathbf{A}(\mathbf{r},t) = \sum\limits_{\tau , \omega,j,m} \left[ a^{\tau}_{\omega j m}\,\mathbf{A}_{\omega j m}^\tau (\mathbf{r},t) +
a^{\tau \, *}_{\omega j m} \,\mathbf{A}_{\omega j m}^{\tau * }(\mathbf{r},t) \right]
\label{VectorPotentialExpansion}
\end{equation}
where the mode functions $\mathbf{A}_{\omega j m}^\tau (\mathbf{r},t)$ have the harmonic time dependence
\begin{equation}
\mathbf{A}_{\omega j m}^\tau (\mathbf{r},t) = \mathbf{A}_{\omega j m}^\tau (\mathbf{r}) \, e^{-i \omega t}
\label{TimeDependenceVectorPotential}
\end{equation}
and $\mathbf{A}_{\omega j m}^\tau (\mathbf{r})$ are given in Eqs.~(\ref{MagneticVectorPotentialKNormalized2}) and (\ref{ElectricVectorPotentialKNormalized2}). The coefficients $a^{\tau}_{\omega j m}$ specify how much of each multipole mode is present in the expansion of the vector
potential. In Eq.~(\ref{VectorPotentialExpansion}), the sums takes\ values $\tau= M$ and $E$, $j=0, 1, 2, \dots$, $m=-j, -j+1, \dots , +j$, and the discrete frequencies $\omega = \omega _{j,n}^\tau$ (where $\tau=M$ or $E$), are given by Eq.~(\ref{AllowedMagneticFrequencies}) and (\ref{AllowedElectricFrequencies}) for magnetic and electric multipoles, respectively. Note that the vector potential $\mathbf{A}(\mathbf{r},t)$ in Eq.~(\ref{VectorPotentialExpansion}) is a real-valued function.

From the classical vector potential mode expansion in Eq.~(\ref{VectorPotentialExpansion}), the quantized vector potential is given by replacing the expansion coefficients with the respective operators~\cite{Davydov_QuantumMechanics}:
\begin{align}
a_{\omega jm}^\tau &\to \hat{a}_{\omega j m}^\tau \nonumber \\
a_{\omega jm}^{\tau \,\, *} &\to \hat{a}_{\omega j m}^{\tau \, \dagger }
\label{CoefficientReplace}
\end{align}
leading to the quantized vector potential operator:
\begin{equation}
\hat{\mathbf{A}}(\mathbf{r},t) = \sum\limits_{\tau ,\omega,j,m} \left[ \hat{a}^{\tau}_{ \omega j m}\,\mathbf{A}_{\omega j m}^\tau (\mathbf{r},t) + \hat{a}^{\tau \, \dagger }_{\omega j m}\,\mathbf{A}_{\omega j m}^{\tau \; * }(\mathbf{r},t) \right]
\label{VectorPotentialOperator}
\end{equation}
where the creation and annihilation operators, $\hat{a}^{\tau \, \dagger }_{\omega j m}$ and $\hat{a}^{\tau }_{\omega j m}$, respectively, satisfy the standard commutation relations~\cite{LL_QuantumElectrodynamics,Davydov_QuantumMechanics}
\begin{equation}
\left[ \hat{a}_{ \omega j m}^\tau, \hat{a}^{\tau^\prime \, \dagger}_{\omega' j' m'} \right] = \delta_{\tau \tau'}\delta_{n n'}\delta_{jj'}\delta_{mm'}
\label{CommuttationRelations0}
\end{equation}
The sum on allowed discrete frequencies $\omega = \omega _{j,n}^\tau$ is the same as in
Eq.~(\ref{VectorPotentialExpansion}). For a given $\{\tau=M, j, m\}$ and $\{\tau=E, j, m\}$, Eqs.~(\ref{AllowedMagneticFrequencies}) and (\ref{AllowedElectricFrequencies}) give the allowed photon frequencies.

Substituting this vector potential operator into Eq.~(\ref{ModeEnergy2}),
and summing the single-mode energies $\mathcal{E}_{\omega jm}^\tau$ over all modes
labeled by $\{\tau,\omega, j, m\}$, leads to the quantized Hamiltonian
\begin{equation}
\hat{H} = \sum\limits_{\tau, n, j, m} \hbar \omega _{j,n}^\tau\left( \hat{a}^{\tau \, \dagger }_{n j m}\,\hat{a}_{n j m}^{\tau} + \frac{1}{2} \right)
\label{Hamiltonian}
\end{equation}
where $\hat{a}^{\tau \, \dagger}_{n j m}$ and $\hat{a}_{n j m}^{\tau}$ are the
creation and destruction operators for electric and magnetic multipole photons, labeled by $\tau=E$ or $\tau=M$, respectively, having energy
$\hbar \omega _{j,n}^\tau$, angular momentum quantum number $j$ and angular momentum projection $m \hbar$ on the z-axis.
To be precise, the modes that are labeled by $\{\tau,\omega, j, m\}$ in Eq.~(\ref{VectorPotentialOperator}) are specified by
$\{\tau, n, j, m \}$ in Eq.~(\ref{Hamiltonian}), where, as stated above, the index $n$ labels the roots ${\omega _{j,n}^\tau}$ sequentially in Eq.~(\ref{AllowedMagneticFrequencies}) and (\ref{AllowedElectricFrequencies}).

The creation and annihilation operators satisfy the commutation relations
\begin{equation}
\left[ \hat{a}_{ n j m}^\tau, \hat{a}^{\tau^\prime \, \dagger}_{n' j' m'} \right] = \delta_{\tau \tau'}\delta_{n n'}\delta_{jj'}\delta_{mm'}
\label{CommuttationRelations}
\end{equation}

The photon number operator is given by
\begin{equation}
\hat{N} = \sum\limits_{\tau njm} \hat{a}^{\tau \dagger }_{njm}\hat{a}^{\tau }_{njm}
\label{PhotonNumberOperator}
\end{equation}

Each field state, or photon, is either an electric multipole, $\tau=E$, or magnetic multipole, $\tau=M$, has the square of angular momentum $j(j+1) \hbar^2$, and z-component of angular momentum $m \hbar$. The parity of the electric photon $\tau = E$ is $(-1)^{j}$ and the parity of the magnetic photon $\tau = M$ is $(-1)^{j+1}$. For each type of photon, electric or magnetic, and for each angular momentum $j$, there is a series of photons of increasing energies labeled by index $n$. As mentioned above, the energies are degenerate with respect to the quantum number $m$.

\section{Allowed Photon Frequencies}
\label{AllowedFrequencies}

The allowed frequencies for magnetic and electric photon multipole modes are given by solving for the roots of Eq.~(\ref{AllowedMagneticFrequencies}) and (\ref{AllowedElectricFrequencies}), respectively, see Tables \ref{Omega-Magnetic} and \ref{Omega-Electric} for several tabulated values. As mentioned above, the allowed photon energies $\hbar \omega$ are degenerate with respect to the quantum number $m$ that specifies the value $J_z$ of angular momentum.
\begin{table}[h!]
\caption{Allowed frequencies for several magnetic multipole photons given by roots of Eq.~(\ref{AllowedMagneticFrequencies}).}
\label{Omega-Magnetic}
\begin{tabular}{c|cccc}
\hline \hline
 $\omega^M_{j,n} R/c$ & n=1 & n=2 & n=3 & n=4 \\ \hline
 j=1 & 4.49341 & 7.72525 & 10.9041 & 17.2208 \\
 j=2 & 5.76346 & 12.3229 & 15.5146 & 18.689 \\
 j=3 & 6.98793 & 10.4171 & 13.698 & 20.1218 \\
 j=4 & 8.18256 & 11.7049 & 15.0397 & 18.3013 \\
\hline \hline
\end{tabular}
\end{table}
\begin{table}[h!]
\caption{Allowed frequencies for several electric multipole photons given by roots of
Eq.~(\ref{AllowedElectricFrequencies}).}
\label{Omega-Electric}
\begin{tabular}{c|cccc}
\hline \hline
$ \omega^E_{j,n} R/c$ & n=1 & n=2 & n=3 & n=4 \\ \hline
 j=1 & 2.74371 & 6.11676 & 9.31662 & 12.4859 \\
 j=2 & 3.87024 & 7.44309 & 10.713 & 13.9205 \\
 j=3 & 4.97342 & 8.72175 & 12.0636 & 15.3136 \\
 j=4 & 6.06195 & 9.96755 & 13.3801 & 16.6742 \\
\hline \hline
\end{tabular}
\end{table}
Note that electric photon frequencies are always smaller than magnetic photon
frequencies for the same $j$ and $n$ values.

\section{Transition Probabilities}
\label{TransitionProbabilities}
It is of some interest to consider the probability of photon absorption or emission by an atom placed inside the spherical cavity. Some general statements can be made about parity selection rules and how the transition matrix elements scale. Exact selection rules based on conservation of parity can be written down. Assume that under inversion of coordinates (under the parity operator), the parity of the initial and final atomic wave functions are $P_i$ and $P_f$, respectively, and the parity of a photon is $P_\gamma$, where parity eigenvalues takes values $\pm 1$. If the initial state of the whole system has one photon, and the final state has no photons, then parity conservation requires $P_i \, P_\gamma = P_f$, or, since each parity eigenvalue is $\pm 1$, parity conservation requires~\cite{LL_QuantumElectrodynamics}
\begin{equation}
P_i \, P_f = P_\gamma
\label{ParityConservation}
\end{equation}
Electric photons, or so-called $Ej$ photons have parity $P^{Ej}_\gamma=(-1)^j$ and magnetic $Mj$ photons have parity $P^{Mj}_\gamma=(-1)^{j+1}$. For absorbing one photon of a given type, parity conservation requires
\begin{equation}
P_i \, P_f = (-1)^j \qquad {\rm electric \,\,\, {\it Ej} \,\,\, photon}
\label{ParityConservationElectric}
\end{equation}
or
\begin{equation}
P_i \, P_f = (-1)^{j+1} \qquad {\rm magnetic \,\,\, {\it Mj} \,\,\, photon}
\label{ParityConservationMagnetic}
\end{equation}

An estimate of the scaling of transition probabilities can be made by considering the atom as a system of charges interacting with the radiation field. Consider an atom in an initial state with energy $E_i$, absorbing a single photon of energy $\hbar \omega $, and making a transition to a final state with energy $E_f$. The transition matrix element contains the quantity $k r = \omega \, r /c$, which enters in the vector potential in Eqs.~(\ref{MagneticVectorPotentialKNormalized})
and~(\ref{ElectricVectorPotentialKNormalized}). The quantity $r$ has a characteristic scale of an atom, on the order of $a\sim 10^{-10}$ m. For example, a typical atomic transition, such as the $3 s \rightarrow 3 p$ transition in Na, has a wavelength $\lambda=$ 590 nm.

Therefore, there is a small dimensionless quantity, $k r \sim k a \sim 2 \pi a / \lambda \sim 0.001$ in the transition matrix elements. The vector potentials in
Eqs.~(\ref{MagneticVectorPotentialKNormalized}) and~(\ref{ElectricVectorPotentialKNormalized}) depend on the spherical Bessel function $j_{l} ( k r )$. Using the small argument, $ kr \ll 1$, series expansion of the Bessel function is
\begin{equation}
j_j(ka) = (ka)^j \left( \frac{2^{ - 1 - j}\sqrt \pi}{\Gamma (j + \frac{3}{2})} + O{(ka)^2} \right)
 \label{SphericalBesselExpansion}
\end{equation}
The vector potential for $Mj$ photons scales as $\mathbf{A}_{kjm}^M (\mathbf{r}) \sim j_j(k\,r)$, and for $Ej$ photons scales as $\mathbf{A}_{kjm}^E ( \mathbf{r} ) \sim
\sqrt{\frac{j+1}{2j+1}} j_{j - 1}( k\,r )$. The probability of absorbing a photon scales as the square of the vector potential (square of the matrix element). Therefore, the ratio of the probability for absorbing an $Mj$ photon, $P(Mj)$, divided by the probability of absorbing an $Ej$ photon, $P(Ej)$, scales as
\begin{equation}
\frac{P(Mj)}{P(Ej)} \sim \frac{(ka)^2}{(j+1)(2j+1)} + O(k a)^4 \sim \frac{10^{-6}}{(j+1)(2j+1)}
\label{ProbMj/EjRatioScaling}
\end{equation}
So the probability for absorption of a magnetic photon is much less than the probability of absorption of an electric photon.

The ratio of the probabilities of absorbing an $E(j+1)$ photon, $P(E(j+1)$, divided by the probability of absorbing an $Ej$ photon, $P(Ej)$, scales as
\begin{equation}
\frac{P(E(j + 1))}{P(Ej)} \sim \frac{j+2}{(j + 1)(2j + 1)(2j + 3)}(ka)^2 + O(ka)^4
\label{ProbE(j+1)/EjRatioScaling}
\end{equation}
From Eq.~(\ref{ProbE(j+1)/EjRatioScaling}), we see that the probability of absorbing higher angular momentum photons rapidly decreases with increasing $j$.

Finally, the ratio of the probability of absorbing an $M(j+1)$ photon to the probability of absorbing an $Mj$ photon scales as
\begin{equation}
\frac{P(M(j + 1))}{P(Mj)} \sim \frac{(ka)^2}{(2j + 3)^2} + O(ka)^{5/2}
\label{ProbM(j+1)/MjRatioScaling}
\end{equation}

\section{Bipartite Entangled Photon States}
\label{BipartiteEntangledPhotonStates}

Two-photon states have been studied in the center-of-momentum frame, and states were enumerated~\cite{Akhiezer-Berestetskii,LL_QuantumElectrodynamics}. Of particular mention is that a two-photon state with total angular momentum state $J=0$ cannot exist. Here, we consider two photon entanglement~\cite{RevModPhys.81.865}. It is well-known that entanglement can be dependent on the frame of reference in which it is observed~\cite{Alsing2002,PhysRevA.68.042102,Lee_2004}. Therefore, in what follows, we assume that the states (wave functions) are describing a system in laboratory coordinates, and not in the center of momentum coordinates. Furthermore, we restrict ourselves to states that are entangled by exchange of a single quantum number, however, the results are somewhat more general, see below. Recently, there has been extensive discussions in the literature about the subtle role of particle indistinguishability in entanglement scenarios~\cite{BENATTI20201,Dieks2020,PhysRevX.10.041012}. Here, our goal is much simpler: to enumerate the possible two-photon entangled states in a sphere in the laboratory coordinate system. We do not explore the complicated matter of whether these states are physically realizable in the laboratory.

\subsection{Entangled Photon Plane Wave Helicity States}
\label{EntangledPhotonHelicityStates}
As a preliminary, consider entanglement of single-photon helicity wave functions defined as~\cite{LL_QuantumElectrodynamics,Varshalovich2021}:
 \begin{equation}
\psi_{\mathbf{p},\lambda}(\mathbf{k},\sigma) = \delta^{(3)}(\mathbf{k} - \mathbf{p})w_\sigma^{(\lambda)}(\mathbf{\hat{p}})
= \langle \mathbf{k},\sigma | \mathbf{p},\lambda \rangle
 \label{PlaneWaveHelicityState}
 \end{equation}
where $w_\sigma ^{(\lambda )}\left( {{\bf{\hat p}}} \right)$ is an eigenfunction of the helicity operator $ ( {{\bf{\hat S}} \cdot {\bf{\hat p}}} )$:
 \begin{equation}
 (\mathbf{\hat{S}} \cdot \mathbf{\hat{p}})w_\sigma^{(\lambda)}(\mathbf{\hat{p}}) = \lambda w_\sigma^{(\lambda)}(\mathbf{\hat{p}})
  \label{HelicityState}
 \end{equation}
Here $\bf{p}$ is the photon momentum along unit vector $\bf{\hat p}$ and $\sigma=+1,0,-1$, is the photon spin coordinate along the z-axis. The full helicity state $\left| {{\bf{p}},\lambda } \right\rangle $ in Eq.~(\ref{PlaneWaveHelicityState}) can be factored into a product of a plane wave state $\left| {\bf{p}} \right\rangle$ and a helicity state $\left| {{\bf{\hat p}},\lambda } \right\rangle$ as:

\begin{equation}
\left| {{\bf{p}},\lambda } \right\rangle = \left| {\bf{p}} \right\rangle \otimes \left| {{\bf{\hat p}},\lambda } \right\rangle
\label{factoredHelicityState}
\end{equation}
where
 \begin{equation}
w_\sigma^{(\lambda)}(\mathbf{\hat{p}}) = \langle \sigma | \mathbf{\hat{p}}\lambda \rangle
\label{HelicityState2}
\end{equation}
The helicity eigenfunction, $w_\sigma ^{(\lambda )}\left( {{\bf{\hat p}}} \right)$, can be written in terms of eigenstates of spin with quantization axis along ${\bf \hat z}$, using the $D$-matrix:
\begin{equation}
 w_\sigma^{(\lambda)}(\mathbf{\hat{p}}) = \langle \sigma | \mathbf{\hat{p}}\lambda \rangle =
 \sum\limits_{\sigma '} D_{\sigma '\lambda}^{(1)}(\mathbf{\hat{p}}) \langle \sigma | \mathbf{\hat{z}},\sigma' \rangle = D_{\sigma \lambda}^{(1)}(\mathbf{\hat{p}})
\label{HelicityWF2}
 \end{equation}
Here $D_{\sigma '\lambda }^{(1)}$ is the Wigner D-matrix for the $j=1$ representation of the 3-d rotation group, that {\it actively} rotates the state
quantized along the z-axis, $\left| {{\bf{\hat z}},\sigma '} \right\rangle $, to the state $\left| {{\bf{\hat p}},\lambda } \right\rangle$, where the spin projection is $\lambda$ along the quantization direction $\bf {\hat{p}}$. Note that $ \langle \sigma | \mathbf{\hat{z}}, \sigma' \rangle = \delta_{\sigma \sigma'} $
because these are the same states.

Associated with the helicity states $\left| {{\bf p},\lambda } \right\rangle$,
we define creation and annihilation operators, $\hat a_{{\bf{k}}\lambda }^\dagger$ and $\hat a_{{\bf{k}}\lambda }$, where $\lambda=+1,-1$, which are the allowed helicity values for photons and $\bf{k}$ is the photon plane wave momentum. The state with $\lambda=0$ is not allowed due to the requirement that the vector potential must be a transverse field.
The creation and destruction operators, $\hat{a}_{{{\bf{k}}}\lambda}^\dagger$ and $\hat{a}_{{{\bf{k}}}\lambda}$, create and destroy photons in states $\left| {{\bf{\hat p}},\lambda } \right\rangle$. These operators satisfy the commutation relations for bosons
\begin{equation}
\left[ \hat{a}_{\mathbf{k}\lambda}, \hat{a}_{\mathbf{k'}\lambda '}^\dagger \right] = \delta_{\mathbf{k},\mathbf{k'}} \delta_{\lambda ,\lambda '}
\label{PlaneWaaveCommutator}
 \end{equation}
The commutation relations ensure that states, and wave functions, created by these operators are symmetric, as required by Bose statistics since photons are spin one bosons.

Entangled bipartite states have been defined for qubit states,
$\left| 0 \right\rangle$ and $\left| 1 \right\rangle$, as~\cite{barnettBook2009}:
\begin{align}
 \left| \Psi^- \right\rangle &= \frac{1}{\sqrt 2} \left( \left| 0 \right\rangle \otimes \left| 1 \right\rangle - \left| 1 \right\rangle \otimes \left| 0 \right\rangle \right) \nonumber \\
 \left| \Psi^+ \right\rangle &= \frac{1}{\sqrt 2} \left( \left| 0 \right\rangle \otimes \left| 1 \right\rangle + \left| 1 \right\rangle \otimes \left| 0 \right\rangle \right) \nonumber \\
 \left| \Phi^- \right\rangle &= \frac{1}{\sqrt 2} \left( \left| 0 \right\rangle \otimes \left| 0 \right\rangle - \left| 1 \right\rangle \otimes \left| 1 \right\rangle \right) \nonumber \\
 \left| \Phi^+ \right\rangle &= \frac{1}{\sqrt 2} \left( \left| 0 \right\rangle \otimes \left| 0 \right\rangle + \left| 1 \right\rangle \otimes \left| 1 \right\rangle \right)
 \label{BellStates}
 \end{align}
These are the iconic entangled two-particle Bell states. We will use these Bell states to factor the two-photon wave functions below. We define four entangled plane wave helicity states for photons:
\begin{align}
 \left| \Psi_1 \right\rangle &= \left( \hat{a}_{\mathbf{k}_1\lambda_1}^\dagger \hat{a}_{\mathbf{k}_2\lambda_2}^\dagger - \hat{a}_{\mathbf{k}_1\lambda_2}^\dagger \hat{a}_{\mathbf{k}_2\lambda_1}^\dagger \right) \left| 0 \right\rangle \nonumber \\
 \left| \Psi_2 \right\rangle &= \left( \hat{a}_{\mathbf{k}_1\lambda_1}^\dagger \hat{a}_{\mathbf{k}_2\lambda_2}^\dagger + \hat{a}_{\mathbf{k}_1\lambda_2}^\dagger \hat{a}_{\mathbf{k}_2\lambda_1}^\dagger \right) \left| 0 \right\rangle \nonumber \\
 \left| \Psi_3 \right\rangle &= \left( \hat{a}_{\mathbf{k}_1\lambda_1}^\dagger \hat{a}_{\mathbf{k}_2\lambda_1}^\dagger + \hat{a}_{\mathbf{k}_1\lambda_2}^\dagger \hat{a}_{\mathbf{k}_2\lambda_2}^\dagger \right) \left| 0 \right\rangle \nonumber \\
 \left| \Psi_4 \right\rangle &= \left( \hat{a}_{\mathbf{k}_1\lambda_1}^\dagger \hat{a}_{\mathbf{k}_2\lambda_1}^\dagger - \hat{a}_{\mathbf{k}_1\lambda_2}^\dagger \hat{a}_{\mathbf{k}_2\lambda_2}^\dagger \right) \left| 0 \right\rangle
 \label{EntangledHelicityStates}
 \end{align}
 where $\lambda_1$ and $\lambda_2$ are two {\it different} helicity values, and $\left| 0 \right\rangle$ is the vacuum state.

For the two-photon helicity states in Eq.~(\ref{EntangledHelicityStates}), wave functions in momentum space can be obtained by projecting these states onto the two-photon helicity basis states:
\begin{equation}
\left| {{\bf{k}}\lambda ,{\bf{k'}}\lambda '} \right\rangle \equiv \hat a_{{\bf{k}}\lambda }^\dagger \hat a_{{\bf{k'}}\lambda '}^\dagger \left| 0 \right\rangle
\label{Two-ParticleBasisStates}
 \end{equation}
The projections of the states in Eq.~(\ref{EntangledHelicityStates}) onto the two-particle basis states in Eq.~(\ref{Two-ParticleBasisStates}) leads to wave functions that can be factored into space $\otimes$ spin wave functions:
\begin{align}
 \langle \mathbf{k}\lambda, \mathbf{k'}\lambda' | \Psi_1 \rangle &= \phantom{-}\Psi_{\mathbf{k}_1,\mathbf{k}_2}^- (\mathbf{k},\mathbf{k'}) \,\, \Psi_{\lambda_1,\lambda_2}^- (\lambda,\lambda') \nonumber \\
 \langle \mathbf{k}\lambda, \mathbf{k'}\lambda' | \Psi_2 \rangle &= -\Psi_{\mathbf{k}_1,\mathbf{k}_2}^+ (\mathbf{k},\mathbf{k'}) \,\, \Psi_{\lambda_1,\lambda_2}^+ (\lambda,\lambda') \nonumber \\
 \langle \mathbf{k}\lambda, \mathbf{k'}\lambda' | \Psi_3 \rangle &= -\Psi_{\mathbf{k}_1,\mathbf{k}_2}^+ (\mathbf{k},\mathbf{k'}) \,\, \Phi_{\lambda_1,\lambda_2}^+ (\lambda,\lambda') \nonumber \\
 \langle \mathbf{k}\lambda, \mathbf{k'}\lambda' | \Psi_4 \rangle &= \phantom{-}\Psi_{\mathbf{k}_1,\mathbf{k}_2}^+ (\mathbf{k},\mathbf{k'}) \,\, \Phi_{\lambda_1,\lambda_2}^- (\lambda,\lambda') \label{FactoredEntangledHelicityStates2}
\end{align}

The wave functions on the left side of Eq.~(\ref{FactoredEntangledHelicityStates2}) have been defined using creation/annihilation operators in Eq.~(\ref{EntangledHelicityStates}), and consequently, these wave functions are all automatically symmetric under exchange of particle coordinates, $\left( {{\bf{k}},\lambda } \right) \leftrightarrow \left( {{\bf{k'}},\lambda '} \right)$, as required by Bose statistics. On the right side of Eq.~(\ref{FactoredEntangledHelicityStates2}), the wave functions $\Psi^+$, $\Phi^+$ and $\Phi^-$ are symmetric under exchange of particle coordinates, and $\Psi^-$ is antisymmetric under exchange of particle coordinates. The wave function $\Psi^-$ is a spin singlet. The wave functions on the right side of Eq.~(\ref{FactoredEntangledHelicityStates2}) have obvious definitions, for example, $\Psi _{{{\bf{k}}_1},{{\bf{k}}_2}}^ - \left( {{\bf{k}},{\bf{k'}}} \right) = \langle \mathbf{k},\mathbf{k'} | \Psi_{\mathbf{k}_1,\mathbf{k}_2}^- \rangle$, where $ | {{\Psi _{{{\bf{k}}_1},{{\bf{k}}_2}}^ - }} \rangle $ is defined in Eq.~(\ref{BellStates}).

\subsection{\label{EntangledSphericalStates}Entangled Photon Spherical States}

Entangled two-photon spherical states can be formed by direct analogy to the entangled two photon plane-wave helicity states in Eq.~(\ref{EntangledHelicityStates}). All that is needed is some quantum number accounting. The entangled states in Eq.~(\ref{EntangledHelicityStates}) are formed by interchanging one quantum number, which takes two different values, $\lambda_1$ and $\lambda_2$. The spherical states have four quantum numbers (instead of two): $ (\tau, \omega, j, m) $.
As in the case of entangled helicity states above, two-photon single-quantum-number entanglement in spherical geometry can be formed by partitioning the quantum numbers into a single entangling quantum number, for example $\omega$, and the rest of the quantum numbers, $(\tau, j, m) $, which we call $\gamma$. There are four ways of creating this partition of the four quantum number $(\tau, \omega, j, m) $:
\begin{align}
\{ \tau | \omega ,j,m \} &= \{ \tau | \gamma \} & \text{Type 1} \label{Partition1} \\
\{ \omega | \tau ,j,m \} &= \{ \omega | \gamma \} & \text{Type 2} \label{Partition2} \\
\{ j | \tau, \omega,m \} &= \{ j | \gamma \} & \text{Type 3} \label{Partition3} \\
\{ m | \tau, \omega,j \} &= \{ m | \gamma \} & \text{Type 4} \label{Partition4}
\end{align}
where $\gamma$ represents the three quantum numbers on the right side of the partition, for each Type of entanglement. Each of the four partitions in
Eq.(\ref{Partition1})-(\ref{Partition4}) represent a type of single-quantum-number bipartite entanglement.

In analogy to Eq.~(\ref{EntangledHelicityStates}), we define the partitions in Eqs.~(\ref{Partition1})-(\ref{Partition4}) as $ \left\{ {\alpha \left| {\gamma} \right.} \right\}$, where the entangling quantum number, $\alpha$, can take one of the values $\alpha= \tau, \omega, j,m$, and $\gamma$ takes the other three values. For example, for Type 2 (frequency) entanglement in Eq.~(\ref{Partition2})
\begin{equation}
\{ \alpha |\gamma \} = \{ \omega |\tau ,j,m \}
\label{SymbolicEntanglement}
\end{equation}
In general, $\alpha$ can take any of the values $\alpha=\tau,\omega, j, m$, and $\gamma$ takes the remaining three values.

For each of the four Types of entanglement in Eqs.~(\ref{Partition1})-(\ref{Partition4}),
there are four cases, in analogy to Eqs.~(\ref{FactoredEntangledHelicityStates2}) for the case of plane wave helicity states. It is a simple matter to replace $\bf{k}$ and $\lambda$ in
Eqs.~(\ref{FactoredEntangledHelicityStates2}) by one of the partitions on the right side of Eqs.~(\ref{Partition1})-(\ref{Partition4}): $(\bf{k},\lambda)\rightarrow (\gamma, \alpha)$, where $\alpha$ takes one of the values $\alpha= \tau, \omega, j, m$, and $\gamma$ takes the remaining three values. Therefore, for each of the Types of entanglement in Eq.~(\ref{Partition1})-(\ref{Partition4}), there are four cases:
\begin{align}
 \langle \gamma\alpha, \gamma'\alpha' | \Psi_1 \rangle &= \phantom{-}\Psi_{\gamma_1,\gamma_2}^- (\gamma,\gamma') \, \Psi_{\alpha_1,\alpha_2}^- (\alpha,\alpha') \label{tangle1} \\
 \langle \gamma\alpha, \gamma'\alpha' | \Psi_2 \rangle &= -\Psi_{\gamma_1,\gamma_2}^+ (\gamma,\gamma') \, \Psi_{\alpha_1,\alpha_2}^+ (\alpha,\alpha') \label{tangle2} \\
 \langle \gamma\alpha, \gamma'\alpha' | \Psi_3 \rangle &= -\Psi_{\gamma_1,\gamma_2}^+ (\gamma,\gamma') \, \Phi_{\alpha_1,\alpha_2}^+ (\alpha,\alpha') \label{tangle3} \\
 \langle \gamma\alpha, \gamma'\alpha' | \Psi_4 \rangle &= \phantom{-}\Psi_{\gamma_1,\gamma_2}^+ (\gamma,\gamma') \, \Phi_{\alpha_1,\alpha_2}^- (\alpha,\alpha') \label{tangle4}
\end{align}
which is just a re-write of Eq.~(\ref{FactoredEntangledHelicityStates2}).
As before, each of the wave functions on the left side of Eqs.~(\ref{tangle1})-(\ref{tangle4}) are symmetric under exchange of the two particle coordinates, since they can be defined in terms of creation/annihilation operators. Therefore, in the spherical case there are sixteen possible types of single-quantum-number entanglement: for each of the four partitions in
Eq.(\ref{Partition1})-(\ref{Partition4}), there are four types of wave functions in Eqs.~(\ref{tangle1})-(\ref{tangle4}).

The meaning of Eqs.~(\ref{tangle1})-(\ref{tangle4}) can be clarified by projecting vector potential modes ${\bf A}^{\tau}_{\omega j m}(\bf{k})$ onto the spherical basis vectors~\cite{LL_QuantumElectrodynamics,Edmonds1960,Varshalovich2021} $ \bf{\hat{e}}_\sigma$ , where $\sigma=+1,0,-1$:
\begin{equation}
\mathbf{A}_{\omega jm}^\tau (\mathbf{k}) \cdot \mathbf{e}_\sigma = A_{\omega jm}^\tau (\mathbf{k},\sigma) = \langle \mathbf{k},\sigma | \tau \omega jm \rangle
\label{SphericalBasisProjection}
\end{equation}
where the vector potential modes are orthonormal:
\begin{equation}
\langle \tau \omega jm | \tau' \omega' j' m' \rangle = \delta_{\tau,\tau'} \, \delta_{\omega,\omega'} \, \delta_{j,j'} \, \delta_{m,m'}
\label{othogonality}
\end{equation}

A comment is in order. For example, for the case of Type 2 (frequency) entanglement in Eq.~(\ref{Partition2}), we can write the entangled state as
\begin{equation}
\left| {{\tau _1}{\omega _1}{j_1}{m_1}} \right\rangle \otimes \left| {{\tau _2}{\omega _2}{j_2}{m_2}} \right\rangle - \left| {{\tau _1}{\omega _2}{j_1}{m_1}} \right\rangle \otimes \left| {{\tau _2}{\omega _1}{j_2}{m_2}} \right\rangle
\label{entangle1}
\end{equation}
where the first ket in the tensor product is associated with the first particle and the second ket with the second particle.
As it stands, Eq.~(\ref{entangle1}) assumes the two photons are distinguishable. This state needs to be symmetrized to obtain a totally symmetric photon state. However, when it is symmetrized, it produces zero. Hence, we needed to choose two different values of $\gamma$, namely $\gamma_1=(\tau_1, j_1, m_1)$ and $\gamma_2=(\tau_2, j_2, m_2)$, so that when the state
\begin{equation}
\left| {{\tau _1}{\omega _1}{j_1}{m_1}} \right\rangle \otimes \left| {{\tau _2}{\omega _2}{j_2}{m_2}} \right\rangle - \left| {{\tau _1}{\omega _2}{j_1}{m_1}} \right\rangle \otimes \left| {{\tau _2}{\omega _1}{j_2}{m_2}} \right\rangle
\label{entangle2}
\end{equation}
is symmetrized, it does not vanish, and so it leads to Eq.~(\ref{tangle2}).

The quantum number partitions in Eq.~(\ref{Partition1})-(\ref{Partition4}) can be made with two (or more) quantum numbers playing the role of $\alpha$, and the remainder of quantum numbers playing the role of $\gamma$. For example, we can choose two quantum numbers: $\alpha=\{\omega,j\}$ and
$\gamma=\{\tau,m\}$. There are 6 combinations taking two quantum numbers at a time (out of 4) to play the role of $\alpha$. This means there are 6 possible partitions, and for each partition, there are 4 entangled symmetric wave functions in Eq.~(\ref{tangle1})-(\ref{tangle4}). Therefore, there are 4+6= 10 different ways to create partitions like in
Eq.~(\ref{SymbolicEntanglement})~\footnote{The case of taking three quantum numbers for $\alpha$ is the same as taking one quantum number for $\alpha$}. For each of those 10 partitions, there are 4 Types of entangled states, given by Eq.~(\ref{tangle1})-(\ref{tangle4}). Consequently, there are 10$\times$ 4 = 40 types of possible entangled states for photons in a spherical cavity.

We have enumerated 40 bipartite entangled photon states in a sphere, however, not all of these states may be physically realizable, for a variety of complicated reasons, which we do not explore here.

\section{Summary}
\label{Summary}
In summary, we have worked out in-detail the iconic problem of photon modes inside a spherical cavity that is bounded by a perfect conductor. Contrary to previous work~\cite{Heitler,Davydov_QuantumMechanics}, we have found that there are two separate conditions for allowed photon frequencies, one condition for magnetic multipole photons and one condition for electric multipole photons, given in Eq.~(\ref{AllowedMagneticFrequencies}) and (\ref{AllowedElectricFrequencies}), respectively. We have written down the quantized vector potential for the computed modes in Eq.~(\ref{VectorPotentialOperator}). In Section \ref{BipartiteEntangledPhotonStates}, we enumerated the bipartite photon states inside the sphere, showing that there are 40 possible bipartite entangled states, however, some of these may not be physically realizable.

\begin{acknowledgments}
The author is grateful to Joseph Bahder for reading the manuscript. The author is thankful to my wife, Margaret Bahder, for her enduring patience during the calculation phase and during the preparation of this manuscript.
\end{acknowledgments}

\appendix

\section{Vector Spherical Harmonics}
\label{VectorSphericalHarmonics}

The vector spherical harmonics, $\mathbf{Y}_{j l m}(\theta ,\phi )$, where $j=l+1, l, |l-1|$ is the total angular momentum, $l$ is the orbital angular momentum and $m$ is the value of component $J_z$ of angular momentum, are defined as direct products of two different irreducible representations of the three-dimensional rotation group, namely $\mathbf{e}_\mu$, and $Y_{lm}(\theta ,\phi ) $, coupled by Clebsch-Gordan coefficients~\cite{Edmonds1960,Davydov_QuantumMechanics,LL_QuantumElectrodynamics,Varshalovich2021}:

\begin{equation}
\mathbf{Y}_{j l m}(\theta ,\phi ) = \sum\limits_\mu \langle 1,l;\mu ,m - \mu | jm \rangle \,\mathbf{e}_\mu \,Y_{l,m - \mu}
\label{YvecDef}
\end{equation}
Here $\mathbf{e}_\mu $ are three spherical basis vectors formed from three spin $S=1$ eigenfunctions, $\chi_\mu (\sigma)$, whose \textit{covariant} components~\cite{Varshalovich2021} are $\mathbf{e}_\mu = \{ \chi_\mu (+1), \chi_\mu (0), \chi_\mu (-1) \}$, where $\sigma= {+1,0,-1}$ is the spin variable~\cite{Davydov_QuantumMechanics,LL_QuantumMechanics}. The $\mathbf{e}_\mu$, $\mu=\{-1,0,+1\}$, are the spherical basis vectors, which are simultaneous eigenfunctions of the spin angular momentum operators $\hat{\mathbf{S}}^2$ and $\hat{S}_z$:
\begin{align}
 \hat{\mathbf{S}}^2 \,\mathbf{e}_\mu &= 2 \mathbf{e}_\mu \nonumber \\
 \hat{S}_z \,\mathbf{e}_\mu &= \mu \mathbf{e}_\mu
\label{SpinEigenfunction}
\end{align}
The covariant spherical basis vectors are related to the Cartesian basis vectors $\{ \mathbf{e}_x, \mathbf{e}_y, \mathbf{e}_z \}$, by~\cite{Edmonds1960,Varshalovich2021}
\begin{align}
 \mathbf{e}_+ &= - \frac{1}{\sqrt{2}}\left( \mathbf{e}_x + i\mathbf{e}_y \right) \nonumber \\
 \mathbf{e}_- &= \frac{1}{\sqrt{2}}\left( \mathbf{e}_x - i\mathbf{e}_y \right) \nonumber \\
 \mathbf{e}_0 &= \mathbf{e}_z
\label{Cartesian-SphericalBasis}
\end{align}
Here, we are using the Condon-Shortley phase convention~\cite{Edmonds1960}, not the convention used by Landau and Lifshitz~\cite{LL_QuantumMechanics,LL_QuantumElectrodynamics}.

In terms of the Cartesian basis vectors, $\{ \mathbf{e}_x, \mathbf{e}_y, \mathbf{e}_z \}$, the action of the spin operator is given by
\begin{equation}
\hat{S}_i \mathbf{e}_k = i \sum_{l=1,2,3} \, e_{ikl} \, \mathbf{e}_l
\label{SpinOpActingOnBasisVector}
\end{equation}
where $e_{ikl}$ is the totally antisymmetric Levi-Civita tensor, where $e_{123}=+1$.

The $Y_{l m }$ in Eq.~(\ref{YvecDef}) are the usual scalar spherical Harmonics that are eigenfunctions of the orbital angular momentum operators $\hat{\mathbf{L}}^2$ and $\hat{L}_z$:
\begin{align}
 \hat{\mathbf{L}}^2 \, Y_{lm} &= l(l+1) Y_{lm} \nonumber \\
 \hat{L}_z \,Y_{lm} &= m Y_{lm}
\label{OrbitalAMeigenfunction}
\end{align}
According to the definitions in Eq.(\ref{YvecDef})--(\ref{OrbitalAMeigenfunction}), the product functions,
$\mathbf{e}_\mu \,Y_{l,m }$, are simultaneous eigenfunctions of $\hat{\mathbf{L}}^2$, $\hat{L}_z$, $\hat{\mathbf{S}}^2$ and $\hat{S}_z$, with eigenvalues $l(l+1)$, $m$, $2$, and $\mu$, respectively.

The spherical harmonic vectors, $\mathbf{Y}_{j l m}(\theta ,\phi )$, form an irreducible representation of the three-dimensional rotation group on the unit sphere, and are simultaneous eigenfunctions of $\hat{J}^2$, $\hat{J}_z$, $\hat{\mathbf{L}}^2$, and $\hat{\mathbf{S}}^2$:
\begin{align}
\hat{J}^2 \mathbf{Y}_{j l m }(\theta, \phi) &= j(j + 1) \mathbf{Y}_{j l m }(\theta, \phi) \label{j2eqn} \\
\hat{J}_z \mathbf{Y}_{j l m }(\theta, \phi) &= m \mathbf{Y}_{j l m }(\theta, \phi) \label{Jzeq} \\
\hat{L}^2 \mathbf{Y}_{j l m }(\theta, \phi) &= l(l + 1) \mathbf{Y}_{j l m }(\theta, \phi) \label{L2eqn} \\
\hat{S}^2 \mathbf{Y}_{j l m }(\theta, \phi) &= 2 \mathbf{Y}_{j l m }(\theta, \phi) \label{S2eqn}
\end{align}
where the total angular momentum operator is $\hat{\mathbf{J}}=\hat{\mathbf{L}} + \hat{\mathbf{S}}$.
The spherical harmonic vectors, $\mathbf{Y}_{j l m }(\theta ,\phi )$, satisfy the orthogonality relation
\begin{equation}
\int d^2\Omega \,\mathbf{Y}_{jlm}^* \left( \theta ,\phi \right) \cdot \mathbf{Y}_{j'l'm'}^{}\left( \theta ,\phi \right) = \delta_{j,\,j'}\,\delta_{l, \,l'}\,\delta_{m, \,m'}
\label{Yorthogonality}
\end{equation}
where the integration is over the unit sphere.

Under the parity operator (spatial inversion), the basis vectors $\mathbf{e}_\mu$ change sign, and we take the scalar spherical harmonics, $Y_{lm}$, to be defined so that they are multiplied by $(-1)^l$, and therefore the spherical harmonic vectors $\mathbf{Y}_{j l m }(\theta ,\phi )$ are eigenfunctions of parity with eigenvalues $(-1)^{l+1}$.
The spherical harmonic vectors $\mathbf{Y}_{jlm}(\hat{\mathbf{r}})$ are not generally transverse to $\hat{\mathbf{r}}$,
where $\hat{\mathbf{r}}=\mathbf{r}/r=(\theta,\phi)$, and so are not appropriate as photon wave functions. However, the functions $\mathbf{Y}_{j j m}(\hat{\mathbf{r}})$ are transverse: $\hat{\mathbf{r}} \cdot \mathbf{Y}_{j j m}(\hat{\mathbf{r}})=0$, and can be taken as possible photon wave functions. These are given the special name "magnetic spherical harmonic vectors": $\mathbf{Y}_{jm}^M(\hat{\mathbf{r}}) \equiv \mathbf{Y}_{jjm}(\hat{\mathbf{r}})$ and have parity $(-1)^{j+1}$. In fact, there are three series of spherical harmonic vectors, defined as~\cite{LL_QuantumElectrodynamics,Davydov_QuantumMechanics}:
\begin{align}
 \mathbf{Y}_{jm}^E &= \frac{1}{\sqrt{j(j + 1)}} \nabla Y_{jm} && P = (-1)^j \label{YE} \\
 \mathbf{Y}_{jm}^M &= \hat{\mathbf{r}} \times \mathbf{Y}_{jm}^E && P = (-1)^{j + 1} \label{YM} \\
 \mathbf{Y}_{jm}^L &= \hat{\mathbf{r}} \, Y_{jm} && P = (-1)^j \label{YL}
\end{align}
where we display the parity $P$ (under transformation of $\mathbf{r} \rightarrow \mathbf{-r}$) of the spherical harmonic vectors~\cite{LL_QuantumElectrodynamics}). The electric spherical harmonic vectors, $ \mathbf{Y}_{jm}^E$, are also transverse, $ \mathbf{Y}_{jm}^E \cdot \hat{\mathbf{r}}=0$, and are possible photon states. However, the spherical harmonic vectors $ \mathbf{Y}_{jm}^L$ are longitudinal and cannot be used for photon wave functions, but they are included in order that they form a complete basis set for expansion of general 3-$d$ vector fields.

The electric and spherical harmonic vectors can also be written as:
\begin{align}
 \mathbf{Y}_{jm}^E &= - i\,\hat{\mathbf{r}} \times \mathbf{Y}_{jm}^M \\
 \mathbf{Y}_{jm}^M &= \frac{1}{\sqrt{j(j + 1)}} \hat{\mathbf{L}} \, Y_{jm} = \mathbf{Y}_{jjm}
 \label{MoreDefinitions}
\end{align}
where the differential angular momentum operator is defined as:
\begin{equation}
\hat{\mathbf{L}} = - i\mathbf{r} \times \nabla
\label{AngularMomentumOperator}
\end{equation}
Some useful identities with the gradient operator and angular momentum are as follows:
\begin{align}
 \mathbf{r} \cdot \hat{\mathbf{L}} &= 0 \\
 \nabla &= \frac{\mathbf{r}}{r}\frac{\partial}{\partial r} - \frac{i}{r^2}\mathbf{r} \times \hat{\mathbf{L}} \\
 \nabla^2 &= \frac{1}{r}\frac{\partial^2}{\partial r^2}r - \frac{\hat{L}^2}{r^2} \\
 i\nabla \times \hat{\mathbf{L}} &= \mathbf{r}\,\nabla^2 - \nabla \left( 1 + r\frac{\partial}{\partial r} \right)
 \label{MoreDefinitions2}
\end{align}
The angular momentum operator can also be defined in terms of the spherical coordinate basis vectors,
 $\{ \hat{\mathbf{r}},\hat{\boldsymbol{\theta}},\hat{\boldsymbol{\phi}} \}, $ as:
 \begin{equation}
 \hat{\mathbf{L}} = - i\hbar \left( \hat{\boldsymbol{\phi}} \frac{\partial}{\partial \theta} - \hat{\boldsymbol{\theta}}\frac{1}{\sin \theta}\frac{\partial}{\partial \phi} \right)
 \label{AngularMomentum}
 \end{equation}

In terms of the angular momentum operator and the scalar spherical harmonics, the spherical harmonic vectors and their parities, $P$, are given by:
\begin{align}
\mathbf{Y}_{jjm}(\hat{\mathbf{r}}) &= \frac{1}{\sqrt{j(j+1)}} \hat{\mathbf{L}} \, Y_{jm}(\hat{\mathbf{r}}) , \quad P = (-1)^{j+1} \nonumber \\
\mathbf{Y}_{j,j-1,m}(\hat{\mathbf{r}}) &= \frac{-1}{\sqrt{j(2j+1)}} \Bigl[ -j \hat{\mathbf{r}} + i \mathbf{r} \times \hat{\mathbf{L}} \Bigr] Y_{jm}(\hat{\mathbf{r}}) , \quad P = (-1)^{j} \nonumber \\
\mathbf{Y}_{j,j+1,m}(\hat{\mathbf{r}}) &= \frac{-1}{\sqrt{(j+1)(2j+1)}} \Bigl[ (j+1) \hat{\mathbf{r}} + i \mathbf{r} \times \hat{\mathbf{L}} \Bigr] Y_{jm}(\hat{\mathbf{r}}) , \quad P = (-1)^{j}
\label{MoreDefs3}
\end{align}

Finally, the electric, magnetic and longitudinal spherical harmonic vectors can be written in terms of spherical harmonic vectors $\mathbf{Y}_{jlm}\left( \hat{\mathbf{r}} \right)$ as:

\begin{align}
 \mathbf{Y}_{jm}^M &= \mathbf{Y}_{jjm} \nonumber \\
 \mathbf{Y}_{jm}^E &= \left( \frac{j}{2j + 1} \right)^{1/2} \mathbf{Y}_{j,j + 1,m} + \left( \frac{j + 1}{2j + 1} \right)^{1/2} \mathbf{Y}_{j,j - 1,m} \nonumber \\
 \mathbf{Y}_{jm}^L &= \left( \frac{j}{2j + 1} \right)^{1/2} \mathbf{Y}_{j,j - 1,m} - \left( \frac{j + 1}{2j + 1} \right)^{1/2} \mathbf{Y}_{j,j + 1,m}
 \label{SHV}
\end{align}

These identities are obtained by using the definitions in Eqs.~(\ref{YvecDef}) and (\ref{YE}) - (\ref{YL}) and using the explicit values for the Clebsch-Gordan coefficients.
These spherical harmonic vectors are orthogonal:
\begin{equation}
\int \mathbf{Y}_{jm}^{\tau \; * } \left( \theta ,\phi \right) \cdot \mathbf{Y}_{j^\prime m^\prime}^{\tau^\prime \;}\left( \theta ,\phi \right)d\Omega = \delta_{\tau ,\tau '}\,\delta_{j,j'}\,\delta_{m,m'}
\label{Yorthogonal}
\end{equation}
where $\tau$ takes the values $\tau=E, M, L$, and $d \Omega = \sin \theta d \theta d \phi$ and the integration is over the unit sphere.

In terms of these spherical coordinate basis vectors, the spherical harmonic vectors in Eq~(\ref{SHV}) are given by:
\begin{align}
 \mathbf{Y}_{lm}^E(\theta ,\phi ) &= \frac{1}{\sqrt{l(l + 1)}} \left( \frac{\partial Y_{lm}}{\partial \theta }\hat{\boldsymbol{\theta}} + \frac{1}{\sin \theta}\frac{\partial Y_{lm}}{\partial \phi }\hat{\boldsymbol{\phi}} \right) \nonumber \\
 \mathbf{Y}_{lm}^M(\theta ,\phi ) &= \frac{- 1}{\sqrt{l(l + 1)}} \left( \frac{1}{\sin \theta}\frac{\partial Y_{lm}}{\partial \phi }\hat{\boldsymbol{\theta}} - \frac{\partial Y_{lm}}{\partial \theta }\hat{\boldsymbol{\phi}} \right) \nonumber \\
 \mathbf{Y}_{lm}^L(\theta ,\phi ) &= \hat{\mathbf{r}}\,Y_{lm}(\theta ,\phi )
 \label{SHV2}
\end{align}

Another notation that is used in the literature for spherical harmonic vectors is~\cite{Barrera1985}:
\begin{align}
 \mathbf{Y}_{lm}^E(\theta ,\phi ) &= \mathbf{\Psi}_{lm}(\theta ,\phi ) \nonumber \\
 \mathbf{Y}_{lm}^M(\theta ,\phi ) &= \mathbf{\Phi}_{lm}(\theta ,\phi ) \nonumber \\
 \mathbf{Y}_{lm}^L(\theta ,\phi ) &= \hat{\mathbf{r}}\,Y_{lm}(\theta ,\phi )
 \label{SHVSphericalBasis}
\end{align}

The spherical harmonic vectors in Eq.~(\ref{SHVSphericalBasis}) form a complete set of vector functions in 3-d space in the sense that an arbitrary well-behaved vector field, $\mathbf{V}(r,\theta,\phi)$, can be expanded as:
\begin{equation}
\mathbf{V}(r,\theta ,\phi ) = \sum\limits_{l = 0}^\infty \sum\limits_{m = - l}^l \left( V_{lm}^r\,\mathbf{Y}_{lm}^L + V_{lm}^E\,\mathbf{Y}_{lm}^E + V_{lm}^M\,\mathbf{Y}_{lm}^M \right)
\label{VectorExpansion}
\end{equation}
and the coefficients, $V_{lm}^r$, $V_{lm}^E $, and $ V_{lm}^M$, can be obtained by using the orthogonality relations for the spherical harmonic vectors in Eq.~(\ref{Yorthogonal}).

\section{Vector Spherical Harmonic Helicity Eigenfunctions}
\label{VectorSphericalHarmonicsHelicityFunctions}

The vector spherical harmonic modes, $\mathbf{A}_{\omega jm}^{M}\left( \mathbf{r} \right)$ and $\mathbf{A}_{\omega jm}^{E}\left( \mathbf{r} \right)$, which are used in expansion of the vector potential in Eqs.~(\ref{VectorPotentialExpansion}) and Eqs.~(\ref{VectorPotentialOperator}), are the Fourier transform of the functions $\mathbf{A}_{\omega jm}^{M}\left( \mathbf{k} \right)$ and $\mathbf{A}_{\omega jm}^{E}\left( \mathbf{k} \right)$, see Eq.~(\ref{FourierTransform}). All relations for spherical harmonic vectors are valid in $\mathbf{k}$-space and in real space. The vector potential modes are eigenfunctions of the parity operator $\hat{P}$:
\begin{align}
 \hat P\mathbf{A}_{\omega j m}^E\left( \mathbf{k} \right) &= (-1)^j \mathbf{A}_{\omega j m}^E\left( \mathbf{k} \right) \nonumber\\
 \hat P\mathbf{A}_{\omega j m}^M\left( \mathbf{k} \right) &= (-1)^{j + 1} \mathbf{A}_{\omega j m}^M\left( \mathbf{k} \right)
\label{VecPotentialParity}
\end{align}
but they are not eigenfunctions of the helicity operator:
\begin{equation}
\hat \Lambda \left( \mathbf{k} \right) = \hat{\mathbf{S}} \cdot \hat{\mathbf{k}}
\label{helicityOp1}
\end{equation}
where $\hat{\mathbf{S}}$ the vector of spin $S=1$ matrices.
However, it is possible to define vector potential modes that are eigenfunctions of the helicity operator~$\hat \Lambda$, but they will not be eigenfunctions of the parity operator. To do so, we define the \textit{vector spherical harmonic helicity eigenfunctions}:
\begin{align}
 \mathbf{Y}_{jm}^{( + 1)}\left( \theta ,\phi \right) &= - \frac{1}{\sqrt{2}}\left( \mathbf{Y}_{jm}^E\left( \theta ,\phi \right) + i\mathbf{Y}_{jm}^M\left( \theta ,\phi \right) \right) \label{VSHHelicity+1} \\
 \mathbf{Y}_{jm}^{( - 1)}\left( \theta ,\phi \right) &= \frac{1}{\sqrt{2}}\left( \mathbf{Y}_{jm}^E\left( \theta ,\phi \right) - i\mathbf{Y}_{jm}^M\left( \theta ,\phi \right) \right) \label{VSHHelicity-1} \\
 \mathbf{Y}_{jm}^{(0)}\left( \theta ,\phi \right) &= \mathbf{Y}_{jm}^L\left( \theta ,\phi \right) = \hat{\mathbf{k}}\,Y_{jm}\left( \theta ,\phi \right)
 \label{VSHHelicity0}
 \end{align}
 for helicity values $\lambda=+1,0,-1$.

As mentioned above, the function $ \mathbf{Y}_{jm}^{(0)}\left( \theta ,\phi \right)$ is longitudinal; it has components in the radial direction $\hat{\mathbf{k}}$ and therefore it is not suitable as the vector potential, which must be transverse in the Coulomb gauge.

These vector spherical harmonic helicity eigenfunctions in Eqs.~(\ref{VSHHelicity+1}) - (\ref{VSHHelicity0}) are orthogonal:
\begin{equation}
\int \mathbf{Y}_{jm}^{(\lambda )\; * } \left( \theta ,\phi \right) \cdot \mathbf{Y}_{j',m'}^{(\lambda')\;}\left( \theta ,\phi \right)d\Omega = \delta_{\lambda ,\lambda '}\,\delta_{j,j'}\,\delta_{m,m'}
\label{HelicityOrtho2}
\end{equation}

Regarding the notation: we use parentheses around helicity eigenvalues, as in $(\lambda)$, and we use no parentheses to label states $\tau=M,E,L$, as in
Eq.~(\ref{VecPotentialParity}).

Equations~(\ref{EigenHelicity}) can be verified by using the identity:
\begin{equation}
( \hat{\mathbf{S}} \cdot \hat{\mathbf{k}} )\mathbf{V} = i\left( \hat{\mathbf{r}} \times \mathbf{V} \right)
\label{HelicityOpIdentity}
\end{equation}
which is valid for any vector function $V$\footnote{The helicity operator can obviously act on functions of position by defining the helicity operator as $\mathbf{S} \cdot \hat{\mathbf{r}}$, where $\hat{\mathbf{r}}$ is the position unit vector. Therefore, all the formulas for the vector spherical harmonic functions and for the helicity operator, can be written as functions of real-space position, or momentum space.}, together with the easily verified identities:
\begin{align}
 ( \hat{\mathbf{S}} \cdot \hat{\mathbf{k}} )\mathbf{Y}_{jm}^E &= i\mathbf{Y}_{jm}^M \label{HelicityOpOnVSH1} \\
 ( \hat{\mathbf{S}} \cdot \hat{\mathbf{k}} )\mathbf{Y}_{jm}^M &= - i\mathbf{Y}_{jm}^E \label{HelicityOpOnVSH2}
 \end{align}
using
\begin{align}
 \mathbf{Y}_{jm}^E ( \hat{\mathbf{k}} ) &= -i \hat{\mathbf{k}} \times \mathbf{Y}_{jm}^M ( \hat{\mathbf{k}} ) \\
 \mathbf{Y}_{jm}^M ( \hat{\mathbf{k}} ) &= \hat{\mathbf{k}} \times \mathbf{Y}_{jm}^E ( \hat{\mathbf{k}} )
 \end{align}
 where $\hat{\mathbf{k}}$ is the unit vector in momentum space defined by the spherical angles $\theta$ and $\phi$, together with the fact that $\hat{\mathbf{k}} \cdot \mathbf{Y}_{j,m}^\tau ( \hat{\mathbf{k}} ) = 0$ for $\tau=M$ or $E$. Therefore, the vector spherical harmonic helicity functions, $\mathbf{Y}_{jm}^{(\lambda )}$, satisfy the eigenfunction equation:
\begin{equation}
( \hat{\mathbf{S}} \cdot \hat{\mathbf{k}} )\mathbf{Y}_{jm}^{(\lambda )} = \lambda \mathbf{Y}_{jm}^{(\lambda )}
\label{helicityVSH}
\end{equation}

Using the definitions of the vector spherical harmonic helicity eigenfunctions, $\mathbf{Y}_{jm}^{(\lambda )} ( \hat{\mathbf{k}} )$, in
Eqs.~(\ref{VSHHelicity+1})-(\ref{VSHHelicity0}), we can define the vector potential \textit{helicty} functions:
\begin{equation}
\mathbf{A}_{\omega jm}^{\left( \lambda \right)}\left( \mathbf{k} \right) = \frac{c}{\omega \sqrt{\delta}} \delta_{k,\omega /c} \mathbf{Y}_{jm}^{(\lambda )} ( \hat{\mathbf{k}} )
\label{VectorPotential-k-Helicity}
\end{equation}
for $\lambda=+1,-1$, which are the analogs of the spherical harmonic vector functions in
Eq.~(\ref{kNormalizedA}),
and where $\mathbf{Y}_{jm}^{(\lambda )} ( \hat{\mathbf{k}} )$ are defined by
Eq.~(\ref{VSHHelicity+1})-(\ref{VSHHelicity0}).
These spherical vector potential modes obviously satisfy the helicity eigenvalue equation~\footnote{As mentioned in the text, we use the notation
$\mathbf{Y}_{jm}^{E}$ or $\mathbf{Y}_{jm}^{M}$ for the electric and magnetic VSH functions, which are written without parenthesis around the $E$ and $M$ labels, as in Eq.~(\ref{kNormalizedA})-(\ref{ElectricVectorPotentialKNormalized}). For VSH \textit{helicity } functions, we use the notation $\mathbf{Y}_{jm}^{(\lambda)}$, with parenthesis around the helicity index $\lambda$.}
\begin{equation}
\hat \Lambda \left( \mathbf{k} \right) \mathbf{A}_{\omega jm}^{\left( \lambda \right)}\left( \mathbf{k} \right) = \lambda \,\mathbf{A}_{\omega jm}^{\left( \lambda \right)}\left( \mathbf{k} \right)
\label{EigenHelicity}
\end{equation}
for $\lambda=+1,0,-1$.

\section{Plane Wave and Spherical Wave Helicity Wave Functions}
\label{HelicityStates}
The angular part of a plane wave helicity wave function is given by~\cite{LL_QuantumElectrodynamics}:
\begin{equation}
\psi_{\hat{\mathbf{p}} \lambda}(\hat{\mathbf{k}}, \sigma)= w^{(\lambda)}_\sigma(\hat{\mathbf{k}}) \,\, \delta^{(2)}(\hat{\mathbf{k}}-\hat{\mathbf{p}})
\label{PlaneWaveHelicityAngular}
\end{equation}
where $\hat{\mathbf{k}}= \mathbf{k}/|\mathbf{k}|$, $\hat{\mathbf{p}}= \mathbf{p}/|\mathbf{p}|$, and $w^{(\lambda)}_\sigma(\hat{\mathbf{p}}) $ are eigenvectors of the helicity operator:
\begin{equation}
(\mathbf{S} \cdot \hat{\mathbf{p}} ) w^{(\lambda)}_\sigma(\hat{\mathbf{p}}) = \lambda w^{(\lambda)}_\sigma(\hat{\mathbf{p}})
\label{HelicityWF1}
\end{equation}
where $\lambda$ is the helicity eigenvalue, which specifies the projection of the spin on the direction of propagation $\hat{\mathbf{p}}$. Note that the delta function is 2-dimensional, in the unit vectors (angles). When integrated over the angles associated with $\hat{\mathbf{k}}$ on the unit sphere, and summed over spins $\sigma$, this wave function is normalized.

As mentioned above, the plane wave helicity wave function in Eq.(\ref{PlaneWaveHelicityAngular}) has no radial part in the wave function. Including a radial part, $\delta(k-p)/k^2$, which specifies the radial momentum (and energy) of the photon, leads to
\begin{equation}
\psi_{\mathbf{p} \lambda}(\mathbf{k}, \sigma)= w^{(\lambda)}_\sigma(\hat{\mathbf{k}}) \delta^{(2)}(\hat{\mathbf{k}}-\hat{\mathbf{p}}) \, \delta(k-p)/k^2
\label{PlaneWaveHelicityWF}
\end{equation}
where $k=\mathbf{k} / |\mathbf{k}|$. The wave function in Eq.~(\ref{PlaneWaveHelicityWF}) is normalized:
\begin{equation}
\sum\limits_{\sigma = - 1}^{ + 1} \int \psi^{\, *}_{\mathbf{p} \lambda}(\mathbf{k},\sigma ) \cdot \psi_{\mathbf{p}^\prime \lambda^\prime }(\mathbf{k},\sigma )\,d^3k =
\delta^{(3)}( \mathbf{p} - \mathbf{p}^\prime) \delta_{\lambda, \lambda^\prime}
\label{PlaneWaveHelicityNorm}
\end{equation}

Note that $\delta^{(2)}(\hat{\mathbf{k}}-\hat{\mathbf{p}}) \, \delta(k-p)/k^2 = \delta^{(3)}(\mathbf{k}-\mathbf{p})$. The wave function in Eq.~(\ref{PlaneWaveHelicityWF}) can be written in terms of the helicity polarization vector
$\mathbf{e}^{(\lambda )}(\hat{\mathbf{k}})$, (rather than the spinor $w^{(\lambda)}_\sigma(\hat{\mathbf{k}})$) and the contravariant spherical basis vector,
$ \mathbf{e}^\sigma$, as:
\begin{equation}
\psi_{\mathbf{p}\lambda }(\mathbf{k},\sigma ) = \delta^{(3)}(\mathbf{k} - \mathbf{p})\,\mathbf{e}^{(\lambda )}(\hat{\mathbf{k}}) \cdot \mathbf{e}^\sigma
\label{NewPlaneWaveHelicityWF}
\end{equation}
where, as usual, the spin variable $\sigma$ in the argument of this wave function is a contravariant index~\cite{LL_QuantumMechanics}.
The polarization vector, $\mathbf{e}^{(\lambda )}(\hat{\mathbf{k}})$, can be written in terms of the $D$-matrix and the covariant~\cite{Varshalovich2021} spherical basis vectors
$\mathbf{e}_{\sigma '}$~~\footnote{In Eq.~(\ref{HelicityVecTransformed}), the index $\lambda$ is really a covariant vector index, so it should be written as a lower index.}:

\begin{eqnarray}
\mathbf{e}^{(\lambda )}(\hat{\mathbf{k}}) = \sum\limits_{\sigma } D_{\sigma' ,\lambda }^{(1)}(\hat{\mathbf{k}}) \,\mathbf{e}_{\sigma'}
\label{HelicityVecTransformed}
\end{eqnarray}
The matrix $D_{\sigma' ,\lambda }^{(1)}(\hat{\mathbf{k}})$ is the rotation matrix for the $j=1$ representation of the 3-$d$ rotation group~\cite{Edmonds1960}.
Using this in Eq.~(\ref{NewPlaneWaveHelicityWF}), together with $\mathbf{e}_{\sigma '} \cdot \mathbf{e}^\sigma = \delta _{\sigma '}^\sigma $, leads to a form for the plane-wave helicity wave function:
\begin{eqnarray}
\psi_{\mathbf{p}\lambda }(\mathbf{k},\sigma ) = D_{\sigma,\lambda }^{(1)}(\hat{\mathbf{k}}) \delta^{(3)}(\mathbf{k} - \mathbf{p})
\label{PlaneWaveHelicity3}
\end{eqnarray}

The wave function $\psi_{\mathbf{p} \lambda}(\mathbf{k}, \sigma)$ is a simultaneous eigenfunction of the momentum operator and helicity operator, but it is not an eigenfunction of the angular momentum operator
$\hat{\mathbf{J}}^2$.

A \textit{spherical wave} helicity function can be defined that is an eigenfunction of the square of the angular momentum, $\hat{\mathbf{J}}^2$, z-component of angular momentum, $\hat{J}_z$, and helicity operator, $\hat{\Lambda}$, as~\cite{LL_QuantumElectrodynamics}
\begin{equation}
\psi_{j m \lambda}(\mathbf{k})= \sqrt{\frac{ 2 j+1}{ 4 \pi}} D^{(j)}_{ \lambda m} (\hat{\mathbf{k}}) \, u^{(\lambda)} ( \hat{\mathbf{k}})
\label{SphericalHelicityWF}
\end{equation}
where $u^{(\lambda)} ( \hat{\mathbf{k}}) = (0, \mathbf{e}^{(\lambda)} ( \hat{\mathbf{k}})) $ is a 4-vector
and $e^{(\lambda)} ( \hat{\mathbf{k}})$ is the the 3-d unit polarization vector perpendicular to $\mathbf{k}$:
\begin{align}
\mathbf{e}^{(\lambda)}(\hat{\mathbf{k}})^* \cdot \mathbf{e}^{(\lambda')}(\hat{\mathbf{k}}) &= \delta_{\lambda,\lambda'} \nonumber \\
\mathbf{k} \cdot \mathbf{e}^{(\lambda')}(\hat{\mathbf{k}}) &= 0
\label{PolarizationVectorAppendix}
\end{align}
and the polarization vectors satisfy the completeness relation:
\begin{equation}
\sum_{\lambda} e_i^{(\lambda)}(\hat{\mathbf{k}}) \, e_j^{(\lambda)}(\hat{\mathbf{k}})^* = \delta_{ij} - \hat{k}_i \hat{k}_j
\label{PolarizationCompleteness}
\end{equation}

In Eq.~(\ref{SphericalHelicityWF}), a particular phase of the Wigner $D$-matrix, $D^{(j)}_{ \lambda m} (\phi,\theta,\gamma)$, is chosen, by defining $D^{(j)}_{ \lambda m} (\hat{\mathbf{k}})=D^{(j)}_{ \lambda m} (\phi,\theta,0)$ where $(\theta,\phi)$ are the spherical polar coordinates associated with $\hat{\mathbf{k}}$. Choosing the $D$-matrix in this way, by taking $\gamma=0$, specifies the phase of the $D$-matrix~\cite{LL_QuantumElectrodynamics}. The normalization of the $D$-matrix in Eq.~(\ref{SphericalHelicityWF}) is then (with $\gamma=0$)~\cite{LL_QuantumElectrodynamics}:
\begin{equation}
\int d\Omega_k D_{\lambda_1,m_1}^{(j_1)} (\hat{\mathbf{k}})^* \cdot D_{\lambda_2,m_2}^{(j_2)}(\hat{\mathbf{k}}) = \frac{4\pi}{2j + 1}\delta_{j_1,j_2}\delta_{\lambda_1,\lambda_2}\delta_{m_1,m_2}
\label{D-matrixNorm}
\end{equation}
The wave function in Eq.~(\ref{SphericalHelicityWF}) is a 4-vector, however, it can be made into a 3-vector by substituting the 3-vector $e^{(\lambda)} ( \hat{\mathbf{k}})$ for the 4-vector $u^{(\lambda)} ( \hat{\mathbf{k}})$.

The signature of the 4-dimensional Minkowskii metric is $(+,-,-,-)$. Therefore, the inner product of the 4-vector $ u^{(\lambda)} ( \hat{\mathbf{k}})^* \cdot u^{(\lambda)} ( \hat{\mathbf{k}})=-1$, leads to a non-intuitive result for the normalization of
$ \psi_{j m \lambda}(\mathbf{k}) $ to be equal to $-1$. To remedy this, we must define a ``physical normalization" of $ u^{(\lambda)} ( \hat{\mathbf{k}})$ to be \textit{defined } as~\cite{LL_QuantumElectrodynamics}
$$ u^{(\lambda)} ( \hat{\mathbf{k}})^* \cdot u^{(\lambda)} ( \hat{\mathbf{k}})= +1$$

As in the case of plane-wave helicity functions, the wave function in Eq.~(\ref{SphericalHelicityWF}) is only the angular part of the complete wave function. Multiplying the wave function in Eq.~(\ref{SphericalHelicityWF} ) by a radial part, $\delta(k-p)/k$ leads to
\begin{equation}
\psi_{pj m \lambda}(\mathbf{k})= \sqrt{\frac{ 2 j+1}{ 4 \pi}} D^{(j)}_{ \lambda m} (\hat{\mathbf{k}}) \, u^{(\lambda)} ( \hat{\mathbf{k}}) \, \delta(k-p)/k
\label{SphericalHelicityWFTotal}
\end{equation}
which results in a normalized wave function:
\begin{equation}
 \int d^3k\, \psi_{pjm\lambda}(\mathbf{k})^* \cdot \psi_{p'j'm'\lambda'}(\mathbf{k}) = \delta(p - p')\,\delta_{jj'}\,\delta_{mm'}\delta_{\lambda \lambda'}
\label{SphericalHelicityWaveFunction}
\end{equation}
where $p$ is the radial momentum.

\section{Expansion of Plane Wave Helicity Functions in terms of Spherical Wave Helicity Functions}
\label{HelicityStatesExpansion}
Just as ordinary scalar plane waves can be expanded in terms of scalar spherical harmonics, $Y_{lm}(\theta,\phi)$, we can expand the plane wave helicity states, $| \mathbf{p}, \lambda \rangle$, given in Eq.~(\ref{PlaneWaveHelicityWF}), in terms of the spherical wave helicity states, $|p^\prime,j,m, \lambda, \rangle$, given by
Eq.~(\ref{SphericalHelicityWFTotal}):
\begin{equation}
| \mathbf{p},\lambda\rangle = \sum_{j=\lambda}^{\infty} \sum_{m=-j}^{+j}
| p^\prime j m \lambda \rangle \langle p^\prime j m \lambda | \mathbf{p}, \lambda \rangle
\label{HelicityExpansion1}
\end{equation}
The sum over $j$ starts at $j=\lambda$ because $j$ can not be less than the projection $\lambda$.
The expansion coefficients are easily found to be:
\begin{equation}
\langle p^\prime j m \lambda^\prime | \mathbf{p}, \lambda \rangle =
\left( \frac{2j + 1}{4\pi} \right)^{1/2} D_{\lambda ,m}^{(j)}(\hat{\mathbf{p}})^* \,\delta_{\lambda ,\lambda '}\frac{\delta (p - p')}{p}
\label{ExpansionCoefficient}
\end{equation}
Note that the $D$-matrix here comes from the definition of the spherical helicity function in Eq.~(\ref{SphericalHelicityWaveFunction}), so there is a special phase convention chosen for the $D$-matrix by taking the third rotation angle $\gamma=0$, see discussion around Eq.~(\ref{D-matrixNorm}).

\section{Transformation of Plane-Wave Helicity Functions under Coordinate Rotation}
\label{RotationofPlaneWaveHelicityWaveFunctions}

Consider Cartesian coordinates $S$, whose basis vectors are $\hat{\mathbf{e}}_k$, and a rotated set of Cartesian coordinates $S^\prime$, with basis vectors $\hat{\mathbf{e}}^\prime_k$, which are related by
\begin{equation}
\mathbf{e'}_i = \sum\limits_{k = 1}^3 R_{ki} \mathbf{e}_k
\label{Rotation}
\end{equation}
We write the plane wave helicity wave function in Eq.~(\ref{PlaneWaveHelicityWF}) as
\begin{equation}
 \psi_{\mathbf{p}\lambda }(\mathbf{k},\sigma ) = \langle \mathbf{k},\sigma | \mathbf{p},\lambda \rangle = \langle \mathbf{k} | \otimes \langle \sigma |\; \cdot \;| \mathbf{p} \rangle \otimes | \hat{\mathbf{p}},\lambda \rangle
\label{PlaneWaveHelicity2}
\end{equation}
where $| \mathbf{p} \rangle$ is the plane wave state and $| \hat{\mathbf{p}},\lambda \rangle$ is the helicity state, given by wave function in Eq.~(\ref{HelicityWF1}). Applying a rotation operator $\hat{R}$, associated with the rotation matrix $R_{ki}$, to the wave function in Eq.~(\ref{PlaneWaveHelicity2}):

\begin{align}
\hat R\psi_{\mathbf{p}\lambda }(\mathbf{k},\sigma ) &\equiv \langle \mathbf{k},\sigma |\hat U(R)|\mathbf{p},\lambda \rangle \nonumber \\
 &= \langle \mathbf{k} | \otimes \langle \sigma |\hat{U}_k(R) \otimes \hat{U}_\sigma (R)| \mathbf{p} \rangle \otimes | \hat{\mathbf{p}},\lambda \rangle \nonumber \\
 &= \langle \mathbf{k}|\hat{U}_k(R)|\mathbf{p} \rangle \langle \sigma |\hat{U}_\sigma (R)|\hat{\mathbf{p}},\lambda \rangle \nonumber \\
 &= \langle \mathbf{k} |\mathbf{R}\,\mathbf{p} \rangle \sum\limits_{\sigma ' = - 1}^{ + 1} \langle \sigma |\hat{U}_\sigma (R)|\sigma ' \rangle \langle \sigma '|\hat{\mathbf{p}},\lambda \rangle \nonumber \\
 &= \delta^{(3)}(\mathbf{k} - \mathbf{R}\,\mathbf{p})\,\sum\limits_{\sigma '} D_{\sigma ,\sigma '}^{(1)}(R) \langle \sigma '|\hat{\mathbf{p}},\lambda \rangle \nonumber \\
\label{RApplied}
\end{align}
where $\hat U(R)$ is the Hilbert space operator associated with the matrix $R_{i k}$. The operator $\hat{U}_k(R)$ acts on the $k$-space and the operator $\hat{U}_\sigma(R)$ acts on spin space, so that $\hat{U}(R)= \hat{U}_k(R) \times \hat{U}_\sigma(R)$.
The terms in Eq.~(\ref{RApplied}) are: Wigner $D$-matrix of the $j=1$ irreducible representation of the rotation group, $D^{(1)}_{\sigma,\sigma^\prime}(R) = \langle \sigma |\hat{U}_\sigma (R)|\sigma ' \rangle$ and the helicity eigenfunction, $w_\sigma ^{(\lambda )}(\hat{\mathbf{p}}) = \langle \sigma |\hat{\mathbf{p}},\lambda \rangle $ in Eq.~(\ref{HelicityWF1}).
The bra-ket $\langle \mathbf{k}|\hat{U}_k(R)|\mathbf{p} \rangle=\langle \mathbf{k} |\mathbf{R}\,\mathbf{p} \rangle=\delta^{(3)}(\mathbf{k} - \mathbf{R}\,\mathbf{p}) $, where $\mathbf{R}$ is the rotation matrix $R_{i k}$, which acts on 3-d Cartesian vectors in Eq.~(\ref{Rotation}). The quantity $\mathbf{R}\,\mathbf{p}$ gives the components of the momentum in the $S^\prime$ coordinates~\footnote{The notation $\mathbf{R}\,\mathbf{p}$ means regular matrix multiplicaton of the components of $\mathbf{R}$ by components of $\mathbf{p}$.}. We use the identity $\delta^{(3)}(\mathbf{k} - \mathbf{R}\,\mathbf{p}) = \delta^{(3)}(\mathbf{R}^{-1}\,\mathbf{k} - \,\mathbf{p})$, which is valid since the Jacobian of the transformation, det $R=1$. Furthermore, we recognize:
$\delta^{(3)}(\mathbf{R}^{-1}\mathbf{k} - \mathbf{p}) = \langle \mathbf{R}^{-1}\mathbf{k} | \mathbf{p} \rangle$.
Equation.~(\ref{RApplied}) becomes
\begin{align}
\hat R \, \psi_{\mathbf{p}\lambda }(\mathbf{k},\sigma )
 &= \langle \mathbf{R}^{-1}\mathbf{k} | \mathbf{p} \rangle \sum\limits_{\sigma '} D_{\sigma ,\sigma '}^{(1)}(R) \langle \sigma '|\hat{\mathbf{p}},\lambda \rangle \nonumber \\
 &= \sum\limits_{\sigma '} D_{\sigma ,\sigma '}^{(1)}(R) \langle \mathbf{R}^{-1}\mathbf{k} | \otimes \langle \sigma '|\; \cdot \;| \mathbf{p} \rangle \otimes | \hat{\mathbf{p}},\lambda \rangle \nonumber \\
 &= \sum\limits_{\sigma '} D_{\sigma ,\sigma '}^{(1)}(R) \langle \mathbf{R}^{-1}\mathbf{k},\sigma ' | \mathbf{p},\lambda \rangle \nonumber \\
 &= \sum\limits_{\sigma '} D_{\sigma ,\sigma '}^{(1)}(R) \,\psi_{\mathbf{p}\lambda }(\mathbf{R}^{-1}\,\mathbf{k},\sigma ') &
\label{TransformedPsi}
\end{align}
where $| \mathbf{p},\lambda \rangle =| \mathbf{p} \rangle \otimes | \hat{\mathbf{p}},\lambda \rangle$ is the plane wave helicity state (which includes the radial part of the wave function). Equation~(\ref{TransformedPsi}) gives the transformed plane wave helicity function, $ \hat{R} \, \psi_{\mathbf{p}\lambda }(\mathbf{k},\sigma )$, under passive rotation of coordinates by rotation operator $\hat R$.

The rotation considered here is a passive rotation, i.e., a rotation of the coordinate system (not a rotation of the states). Both Refs.~\cite{Edmonds1960} and \cite{LL_QuantumMechanics,LL_QuantumElectrodynamics} use the same convention of passive rotations, and they use the same conventions for defining the Wigner $D$-matrix. See the excellent discussions of rotation operators in Refs.~\cite{Wolf1969,Morrison1987,Mann2016}.

\section{Transformation of Spherical-Wave Helicity Functions under Coordinate Rotations}
\label{RotationsSphericalHelicityWaveFunctions}

Consider a rotation $R$ of Cartesian coordinates $S$, with basis vectors $\hat{\mathbf{e}}_k$, to Cartesian coordinates $S^\prime$ with basis vectors $\hat{\mathbf{e}}^\prime_k$, given by
Eq.~(\ref{Rotation}). This is a passive rotation~\cite{LL_QuantumMechanics,Morrison1987,Mann2016}. We take the inner product of the spherical-wave helicity states, in Eq.~(\ref{SphericalHelicityWFTotal}), with the Cartesian basis vectors $\mathbf{e}_n$~\footnote{By taking the inner product with Catesian basis vectors, we are computing the Cartesian components of the spherical-wave helicity function. This function contains the 4-vector $u^{(\lambda)}(\hat{\mathbf{k}})$, which can be interpreted as the 3-vector $e^{(\lambda)}(\hat{\mathbf{k}})$, in order to facilitate the inner product.}:
\begin{align}
\psi_{pjm\lambda }(\mathbf{k},n) &= \langle \mathbf{k},n|pjm\lambda \rangle \nonumber \\
 &= \left( \frac{2j + 1}{4\pi} \right)^{1/2} D_{\lambda ,m}^{(j)}(\hat{\mathbf{k}})\, \mathbf{e}^{(\lambda )}(\hat{\mathbf{k}}) \cdot \mathbf{e}_n
\label{SphericalHelicity2}
\end{align}

The rotated spherical-wave helicity function is:
\begin{align}
\hat{R} \psi_{pjm\lambda}(\mathbf{k},n) &= \langle \mathbf{k},n | \hat{U}(R) | pjm\lambda \rangle \nonumber \\
&= \sum_{n' = \{x,y,z\}} \int d^3k' \, \langle \mathbf{k},n | \hat{U}(R) | \mathbf{k}',n' \rangle \langle \mathbf{k}',n' | pjm\lambda \rangle
\label{RotationSphericalHelicity3}
\end{align}
where we inserted a complete set of plane wave states $ | \mathbf{k}^\prime, n^\prime\rangle$. Writing the rotation operator as a product of operators that acts in $k$-space and on Cartesian basis vectors $\hat{U}( R)=\hat{U}_k \times \hat{U}_n$, Eq.~(\ref{RotationSphericalHelicity3}) becomes
\begin{align}
 \hat R\, \psi_{pjm\lambda }(\mathbf{k},n) &= \sum\limits_{n' } \int d^3k'\, \langle \mathbf{k}|\hat{U}_k|\mathbf{k'} \rangle \langle n|\hat{U}_n|n' \rangle \langle \mathbf{k'},n'|pjm\lambda \rangle \nonumber \\
 &= \sum\limits_{n' } \int d^3k'\, \delta^{(3)}(\mathbf{k} - \hat R\mathbf{k'}) R_{n,n'} \langle \mathbf{k'},n'|pjm\lambda \rangle
\label{RotationSphericalHelicity4}
\end{align}
where $R_{n,n'}$ is the rotation matrix that transforms the Cartesian basis vectors, given in Eq.~(\ref{Rotation}). Using the identity $\delta^{(3)}(\mathbf{k} - \mathbf{R}\,\mathbf{k'}) = \delta^{(3)}(\mathbf{k'} - \mathbf{R}^{-1}\,\mathbf{k})$, and perfoming the integral over $d^3 k^\prime$, we obtain the rotated spherical helicity wave function:
\begin{equation}
\hat R\,\psi_{pjm\lambda }(\mathbf{k},n) = \sum\limits_{n' = \{ x,y,z\} } R_{n n'}\, \psi_{pjm\lambda }(R^{-1}\,\mathbf{k},n')
\label{RotationSphericalHelicity5}
\end{equation}
In Eq.~(\ref{RotationSphericalHelicity5}), the index $n$ labels the Cartesian components of the wave function.

Alternatively, the spherical helicity function can be written in terms of the 4-vector amplitude $u^{(\lambda)}$ as in Eq.~(\ref{SphericalHelicityWFTotal}). It can also be written in terms of
the 3-d helicity vector $ \mathbf{e}^{(\lambda)}(\hat{\mathbf{k}}) $:
\begin{equation}
\psi_{j m \lambda}(\mathbf{k})= \sqrt{\frac{ 2 j+1}{ 4 \pi}} D^{(j)}_{ \lambda m} (\hat{\mathbf{k}}) \, \mathbf{e}^{(\lambda)} ( \hat{\mathbf{k}}) \, \delta(k-p)/k
\label{SphericalHelicityWFTotalPolVector}
\end{equation}
Again, consider a (passive) rotation of coordinate system with operator $\hat R$. Under this rotation, the wave vector transforms as $\mathbf{k}^\prime= \hat R \, \mathbf{k}$. Under this same transformation, $\hat R$, the kets transform as
\begin{equation}
|\mathbf{k}^\prime\rangle = \hat{U}( R) |\mathbf{k}\rangle
\label{KetTransform}
\end{equation}
Under this same transformation, the spherical helicity wave function
in Eq.~(\ref{SphericalHelicityWFTotalPolVector}) transforms as:
\begin{align}
 \hat R\,\psi_{pjm\lambda }(\mathbf{k}) &= \langle \mathbf{k}|\hat R|pjm\lambda \rangle \nonumber \\
 &= \langle \mathbf{k} | \sum\limits_{m'} | pjm'\lambda \rangle \langle pjm'\lambda |\hat R|pjm\lambda \rangle \nonumber \\
 &= \sum\limits_{m'} \langle \mathbf{k}|pjm'\lambda \rangle D_{m',m}^{(j)}(R) \nonumber \\
 &= \sum\limits_{m'} \psi_{pjm'\lambda }(\mathbf{k}) \,D_{m',m}^{(j)}(R)
\label{RotateK}
\end{align}
We have used the fact that the wave functions, $\psi_{pjm\lambda }(\mathbf{k})$, transform according to
the $j$th irreducible representation of the rotation group, i.e., the functions transform with the $D_{m',m}^{(j)}(R)$ matrix.
We have that the wave vector transforms according to $|\mathbf{k} \rangle = \hat{R}^{-1}\,|\mathbf{k'} \rangle $
and taking the Hermitian conjugate
\begin{equation}
\langle \mathbf{k} | = \langle \mathbf{k'} | (\hat{R}^{-1})^\dagger = \langle \mathbf{k'} |\hat R = \langle R^{-1}\,\mathbf{k'} |
\label{HermConjug}
\end{equation}
Using this in Eq.~(\ref{RotateK}) leads to the transformation of the spherical helicity wave function

\begin{equation}
\hat R\,\psi_{pjm\lambda }(\mathbf{k},\sigma ) = \sum\limits_{m'} \psi_{pjm'\lambda }(R^{-1}\,\mathbf{k'}) \,D_{m',m}^{(j)}(R)
\label{HelicityWFTransformation}
\end{equation}
Equation~(\ref{HelicityWFTransformation}) is exactly the transformation we would expect for functions that are a basis for the $j$th irreducible representation of the 3-$d$ rotatoin group.

To emphasize again, the transformation operator, $\hat R$, in Eq.~(\ref{HelicityWFTransformation}) can be an active or passive transformation. Of course, the matrix elements of $D_{m',m}^{(j)}(R)$ have to be chosen for the type of rotation being performed.

\section{Note on Rotation of States and Vectors with the $D$-Matrix}
\label{RotationsUsingD-Matrix}

Consider a vector $\mathbf{V}$:
\begin{equation}
\mathbf{V} = \sum\limits_i V^i \mathbf{e}_i = \sum\limits_i V_i \mathbf{e}^i =
 \sum\limits_k \bar{V}_k \bar{\mathbf{e}}_k
\label{vector}
\end{equation}
The vector $\mathbf{V}$ can be expressed in terms of \textit{covariant} basis vectors $\mathbf{e}_i$, or contravariant basis vectors, $\mathbf{e}^i$~\cite{Varshalovich2021}. The covariant and contravariant basis vectors satisfy $ \mathbf{e}_i \cdot \mathbf{e}^j= \delta_{i}^{j}$.

The covariant and contravariant basis vectors for Cartesian coordinates are the same,
$\mathbf{e}^k = \mathbf{e}_k \equiv \bar{\mathbf{e}}_k$, for $k=\{x, y, z\}$ as are the Cartesian components of a vector $V^k=V_k=\bar{V}_k$~\footnote{Here, we speak about passive rotations, where the coordinate system is rotated with respect to the initial coordinates $S$. Alternatively, we could carry out active rotations, where the rotation operator $\hat R$ rotates the vector, keeping the coordinates fixed. In what follows, all formulas are correct for both the passive and active rotattions, only the form of $\hat R$ is different.}. In general, basis vectors are not orthogonal, so there are two possible basis vectors, as in Eq.(\ref{vector}). A general vector $\mathbf{V}$ can also be expressed in terms of the ``spherical basis" vectors
\begin{align}
 \mathbf{e}_{+1} &= - \frac{1}{\sqrt{2}}\left( \bar{\mathbf{e}}_x + i\bar{\mathbf{e}}_y \right) \nonumber \\
 \mathbf{e}_0 &= \bar{\mathbf{e}}_z \nonumber \\
 \mathbf{e}_{-1} &= \frac{1}{\sqrt{2}}\left( \bar{\mathbf{e}}_x - i\bar{\mathbf{e}}_y \right)
\label{transformmation}
\end{align}
where $\bar{\mathbf{e}}_x$, $\bar{\mathbf{e}}_y$, and $\bar{\mathbf{e}}_z$, are Cartesian basis vectors. The new basis vectors, $\mathbf{e}_\sigma$, for $\sigma=+1, 0, -1$, are called the covariant ``spherical basis" vectors~\footnote{We are using the Condon-Shortley phase convention here for the basis vector~\cite{Edmonds1960}. Note that Landau and Lifshitz~\cite{LL_QuantumMechanics,LL_QuantumElectrodynamics} use a different phase convention for the basis vectors. Landau and Lifshitz~\cite{LL_QuantumMechanics,LL_QuantumElectrodynamics} use passive rotation of coordinates to define their $D$-matrix. Landau and Lifshitz use the same $D$-matrix as is used by Edmonds~\cite{Edmonds1960}, who also uses passive rotations in his Ref.~\cite{Edmonds1960}. Previous work by Edmonds used different conventions.}. The spherical coordinate system associated with the basis vectors $\mathbf{e}_\sigma$ (and $\mathbf{e}^\sigma$) is not orthogonal, so the covariant and contravariant basis vectors are not the same. One can freely use the $D$-matrix definitions tablulated in Landau and Lifshitz~\cite{LL_QuantumMechanics,LL_QuantumElectrodynamics} , and in Edmonds 1960 edition of his book~\cite{LL_QuantumMechanics,LL_QuantumElectrodynamics}, to perform rotations of regular 3-$d$ vectors, as follows.

Consider a Cartesian coordinate system $S$, and a vector $\mathbf{V}$ with Cartesian components
$V_k $ in $S$. In a rotated system of coordinates, $S^\prime$, this same vector has components $V_k^\prime$. The vector components in $S$ can be related to the vector components in $S^\prime$ by the rotation matrix $R_{i k}$:
\begin{equation}
{V_i }^\prime = \sum\limits_\beta R_{i k }\,V_k
\label{RotationOp}
\end{equation}
Here, the rotation matrix elements are defined as matrix elements between Cartesian basis vectors:
\begin{equation}
R_{ik} = \langle \bar{\mathbf{e}}_i|\hat R(\alpha ,\beta ,\gamma )|\bar{\mathbf{e}}_k \rangle
\label{RotattionMatrix}
\end{equation}
where $\hat{R}(\alpha ,\beta ,\gamma )$ is the rotation operator that is specified in terms of Euler angles $\{\alpha ,\beta ,\gamma\}$~\cite{Edmonds1960,LL_QuantumMechanics,Varshalovich2021}.

Alternatively, the vector $\mathbf{V}$ can be expressed in spherical components as:
\begin{equation}
\mathbf{V} = \sum\limits_{\sigma = - 1}^{ + 1} V_\sigma \mathbf{e}^\sigma
\label{SphericalComponents}
\end{equation}
Note that $V_\sigma$ are the \textit{covariant} spherical components of vector $\mathbf{V}$.
Under the rotation the vector spherical components in coordinate system $S$ and $S^\prime$ are related by the Wigner $D$-matrix $D^{(j)}_{\sigma^\prime, \sigma}$ for the $j=1$ irreducible representation of the 3-$d$ rotation group. The Wigner $D$-matrix is designed to rotate \textit{covariant} components of vectors, and it is also designed to rotate \textit{covariant} basis vectors $\mathbf{e}_\sigma$. The caveat is that, in both cases, they must be the covariant components~\cite{Varshalovich2021}. Both the \textit{covariant} components of vectors, $V_\sigma$, and the covariant basis vectors, $\mathbf{e}_\sigma$, transform under rotations according to the $j=1$ representation of the 3-d rotation group. The components of the vector in coordinates $S$ and $S^\prime$ are related by:
\begin{equation}
{V_\sigma }^\prime = \sum\limits_{\sigma^\prime = - 1}^{ + 1} D_{\sigma ',\sigma }^{(1)}(\alpha ,\beta ,\gamma )\,V_{\sigma '} =D_{\sigma,\sigma^\prime }^{(1)}(\alpha ,\beta ,\gamma )^{\, *}\,V_{\sigma '}
\label{D-matrixRotation}
\end{equation}
where $V_{\sigma^\prime}$ are the spherical vector components in the $S$ coordinate system and $V^{\prime}_{\sigma} $ are the spherical vector components in the $S^\prime$ system. Note the position of the $\sigma$ and $\sigma^\prime$ indices. Basis vectors $\mathbf{e}_\sigma$ transform in the same way as the components $V_\sigma$. If we are given vector components that are not covariant spherical components, then we need to transform the components to spherical components before Eq.~(\ref{D-matrixRotation}) is applied~\cite{Varshalovich2021}. For example, if we know the Cartesian vector components $\bar{V}_k$, where $k=x,y,z$, the spherical components, $V_\sigma$, where $\sigma=+1,0,-1$, are given by~\cite{Edmonds1960,LL_QuantumMechanics,LL_QuantumElectrodynamics}
\begin{align}
 V_{+ 1} &= - \frac{1}{\sqrt{2}}\left( \bar{V}_x + i\,\bar{V}_y \right) \nonumber \\
 V_0 &= \bar{V}_z \nonumber \\
 V_{- 1} &= \frac{1}{\sqrt{2}}\left( \bar{V}_x - i\bar{V}_y \right)
\label{VecTransformation}
\end{align}
Equation~(\ref{D-matrixRotation}) can then be applied to compute the vector components $V_\sigma^\prime$ in the $S^\prime$ system. Note that the vector components in Eq.~(\ref{VecTransformation}) transform from Cartesian to spherical form in the same way as the basis vectors in Eq.~(\ref{transformmation}). The covariant spherical vector components can also be obtained by taking the inner product of the vector $\mathbf{V}$ with the covariant spherical basis vectors:
\begin{equation}
V_\sigma = \mathbf{V} \cdot \mathbf{e}_\sigma
\label{DotProductComponents}
\end{equation}

In general, the $D$-matrix can be used to actively or passively rotate states that have definite angular momentum, $j$, and definite z-component of angular momentum, $j_z$:
\begin{equation}
| jm \rangle = \sum\limits_{m^\prime = -j}^{ +j} D_{m',m}^{(j)}(\alpha ,\beta ,\gamma )\,| jm' \rangle \,
\label{JMRotation}
\end{equation}
where $| jm \rangle$ are states labeled by irreducible representation $j$ of the 3-d rotation group, and $j$ taking integer or half-integer values, and as usual $m=-j,-j+1,\cdots,+j$. An example of such states are the spherical harmonics, $| l m \rangle =Y_{l m}$, which have integer angular momentum $j=l$. An example of using Eq.~(\ref{JMRotation}) to transform states is given in Appendix~\ref{RotationofPlaneWaveHelicityWaveFunctions}.

\textit{Example: Transformation of a vector under passive rotation of coordinates using the $D$-matrix.} A vector with Cartesian components can be transformed by the rotation matrix $R_{i k}$ in
Eq.~(\ref{RotationOp}). However, it is instructive to carry out the transformation using the Wigner $D$-matrix, as discussed above.
Define a vector $\mathbf{A}$ by its Cartesian coordinates in the coordinate system $S$:
\begin{equation}
\mathbf{A}= (A_x,A_y,A_z)=(1,0,0)
\label{VectorToTransForm}
\end{equation}
This vector is along the $x$-axis in coordinate system $S$. Now rotate the $\{x,y,z\}$ coordinate system by angle $\beta=\pi/2$ around the $y$-axis. The new $z^\prime$-axis becomes the old $x$-axis. Obviously, the new components of the same vector $\mathbf{A}$ in the new (rotated) Cartesian $S^\prime$ coordinates are
\begin{equation}
\mathbf{A}= (A^\prime_x, A^\prime_y, A^\prime_z)=(0,0,1)
\label{RotatedVectorComponents}
\end{equation}

Using the Wigner $D$-matrix, we can do the same rotation. The $D$-matrix is defined by three successive rotations $\{\alpha, \beta,\gamma \}$ of the coordinate system, first rotating by angle $\alpha$ around the $z$-axis, then rotating by angle $\beta$ around the \textit{new} $y$-axis, and finally rotating by angle $\gamma$ around the \textit{new} $z$-axis. The $D$-matrix for passive rotations is defined as~\cite{Edmonds1960,LL_QuantumMechanics}:
\begin{equation}
D_{m',m}^{(j)}(\alpha ,\beta ,\gamma ) = e^{im'\gamma }\,d_{m',m}^j(\alpha ,\beta ,\gamma )\,e^{im\alpha }
\label{DMatrix2}
\end{equation}
For the rotation of coordinate system $S$ by angle $\beta=\pi/2$ around the $y$-axis, the angles $\alpha=\gamma=0$. The spherical components of the vector $\mathbf{A}$ are transformed under rotations by the matrix $D_{m',m}^{(1)}(0,\pi /2,0)$ for the $j=1$ representation. The $D$-matrix for this passive rotation is~\cite{Edmonds1960,LL_QuantumMechanics}:
\begin{align}
D_{m',m}^{(j)}(0,\pi /2,0) &= \,d_{m',m}^j(0,\pi /2,0)\, \nonumber \\
 &= \begin{pmatrix}
 1/2 & 1/\sqrt{2} & 1/2 \\
 - 1/\sqrt{2} & 0 & 1/\sqrt{2} \\
 1/2 & - 1/\sqrt{2} & 1/2
 \end{pmatrix}
\label{DmatrixDefinition}
\end{align}
The $D$-matrix takes as input spherical components of a vector.
Using the transformation from Cartesian to spherical components in Eq.~(\ref{VecTransformation}), the spherical components of vector $\mathbf{A}$ in the coordinate system $S$ are:
\begin{equation}
A_{+1}=-1/\sqrt{2}, \quad A_{0}=0, \quad A_{-1}=1/\sqrt{2}
\label{SphericalComponentsVector}
\end{equation}
Using the transformation of spherical components in Eq.~(\ref{D-matrixRotation}), together with the specific $D$-matrix in Eq.~(\ref{DmatrixDefinition}), the new spherical components of vector $\mathbf{A}$ in the $S^\prime$ system are:
\begin{equation}
\begin{pmatrix}
 A'_{ + 1} \\
 A'_0 \\
 A'_{ - 1}
\end{pmatrix} = \begin{pmatrix}
 1/2 & 1/\sqrt{2} & 1/2 \\
 - 1/\sqrt{2} & 0 & 1/\sqrt{2} \\
 1/2 & - 1/\sqrt{2} & 1/2
 \end{pmatrix} \begin{pmatrix}
 - 1/\sqrt{2} \\
 0 \\
 1/\sqrt{2}
 \end{pmatrix} \, = \begin{pmatrix}
 0 \\
 1 \\
 0
 \end{pmatrix}
\label{MatrixMultiplly}
\end{equation}
Using Eqs.~(\ref{VecTransformation}) to transform the new spherical components, $ ({{A'}_{ + 1}},{{A'}_0},{{A'}_{ - 1}}) $, back to the new Cartesian components $(A_x^\prime,A_y^\prime,A_z^\prime)$ in system $S^\prime$, leads to the components of vector $\mathbf{A}$ in Eq.~(\ref{RotatedVectorComponents}), as expected by intuiton.

\section{Some Tabulated Integrals and Recurrence Relations}
\label{UsefulIntegrals}

Some useful integrals are listed below for reference.

\begin{equation}
\int Y_{jm}(\hat{\mathbf{k}}) \, e^{i \mathbf{k} \cdot \mathbf{r}} \, d\Omega_k = g_j(kr) \, Y_{jm}(\hat{\mathbf{r}})
\label{ScalarYFT}
\end{equation}
where the $g_l(k r)$ function is given by:
\begin{equation}
g_l(k r)=4 \pi i^l j_l(k r)
\label{g-function}
\end{equation}
where $j_l(k r)$ is the spherical Bessel function of order $l$.
\begin{equation}
\int \mathbf{Y}_{jlm}(\hat{\mathbf{k}}) \,e^{i\mathbf{k} \cdot \mathbf{r}}\,d\Omega_k = g_l(kr)\mathbf{Y}_{jlm}(\hat{\mathbf{r}})
\label{FourierIntegral1}
\end{equation}
where $\mathbf{Y}_{j l m}(\hat{\mathbf{k}})$ is given by Eq.~(\ref{YvecDef}) and $ d\Omega_k=\sin\theta d\theta d \phi $.

For $\mathbf{Y}_{jm}^M(\hat{\mathbf{k}})$ and $\mathbf{Y}_{jm}^E(\hat{\mathbf{k}})$ in Eq.~(\ref{YM}) and (\ref{YL}), their Fourier transforms are given by:
\begin{equation}
\int \mathbf{Y}_{jm}^M(\hat{\mathbf{k}}) \,e^{i\mathbf{k} \cdot \mathbf{r}}\,d\Omega_k =
g_j(kr)\mathbf{Y}_{jm}^M(\hat{\mathbf{r}})
\label{FourierIntegral2}
\end{equation}
\begin{align}
 \int & \mathbf{Y}_{jm}^E (\hat{\mathbf{k}}) \, e^{i\mathbf{k} \cdot \mathbf{r}} \, d\Omega_k = \nonumber \\
 & \quad \frac{1}{\sqrt{2j + 1}} \left[ \sqrt{j} \, g_{j + 1}(kr) \mathbf{Y}_{j,j + 1,m}(\hat{\mathbf{r}}) + \right. \nonumber \\
 & \qquad \left. \sqrt{j + 1} \, g_{j - 1}(kr) \mathbf{Y}_{j,j - 1,m}(\hat{\mathbf{r}}) \right]
\end{align}

The integral of Bessel functions:
\begin{equation}
\int\limits_0^1 x\,J_\nu (\alpha x) \,J_\nu (\beta x)\,dx = \begin{cases}
 0, & \alpha \ne \beta \\
 \frac{1}{2} [J_{\nu + 1}(\alpha )]^2, & \alpha = \beta
 \end{cases}
\label{BesselIntegral}
\end{equation}
where $\alpha$ and $\beta$ are roots: $J_\nu(\alpha)=J_\nu(\beta)=0$, and $\nu>-1$.

The spherical Bessel functions satisfy the recurrence relations:
\begin{equation}
j'_l(x) = \frac{l}{x} \, \,j_l(x) - j_{l + 1}(x)
\label{jIdentity1}
\end{equation}

\begin{equation}
j'_l(x) = j_{l - 1}(x) - \frac{l + 1}{x} \,\, j_l(x)
\label{jIdentity2}
\end{equation}

Equation~(\ref{sphericalBesselFn}) in the text gives the relation between the spherical Bessel functions, $j_l(x)$, and the ordinary Bessel functions of the first kind, $J_\nu(x)$, see also Ref.~\cite{Jackson-3rdEdition}.

Finally, for completeness, we include the expansion of a scalar plane wave in terms of (scalar) spherical harmonics:
\begin{equation}
e^{i\mathbf{k} \cdot \mathbf{r}} = 4\pi \sum\limits_{l = 0}^\infty \sum\limits_{m = - l}^l i^l\,j_l(kr) \,Y_{lm}(\hat{\mathbf{k}})^*\;Y_{lm}(\hat{\mathbf{r}})
\label{PlaneWaveExpansion}
\end{equation}
The analogous expression for the expansion of a (vector) plane-wave helicity state in terms of spherical-wave helicity states is given in the text in Eq.~(\ref{HelicityExpansion1}).

%
\bibliographystyle{apsrev4-2}
\bibliography{References-Quantum}
%
\end{document}